\documentclass[12pt]{amsart}

\usepackage[utf8]{inputenc}
\usepackage[english]{babel}
\usepackage[margin=2.5cm]{geometry}	
\usepackage{color}
\usepackage{xcolor}
\usepackage{enumerate}
\usepackage{enumitem}
\usepackage{amsmath}
\usepackage{amsthm}
\usepackage{amssymb}
\usepackage{dsfont}

\newcommand{\sub}[1]{_{\mathrm{#1}}}

\newcommand{\Id}{\mathds{1}}
\newcommand{\id}{\mathbb{I}}
\newcommand{\TRS}{\Theta}

\newcommand{\N}{\mathbb{N}}
\newcommand{\Z}{\mathbb{Z}}

\newcommand{\R}{\mathbb{R}}
\newcommand{\C}{\mathbb{C}}
\newcommand{\T}{\mathbb{T}}

\newcommand{\K}{\mathcal{K}}
\newcommand{\Hi}{\mathcal{H}}
\newcommand{\cH}{\mathcal{H}}

\newcommand{\Hf}{\mathcal{H}\sub{f}}
\newcommand{\U}{\mathcal{U}}
\newcommand{\rU}{\mathrm{U}}

\newcommand{\rSp}{\mathrm{Sp}}

\newcommand{\UZ}{\mathcal{U}\sub{BFZ}}

\newcommand{\F}{\mathcal{F}}

\newcommand{\inn}[2]{\left\langle #1, #2 \right\rangle}

\DeclareMathOperator{\tr}{tr}
\DeclareMathOperator{\Span}{Span} 
\DeclareMathOperator{\diag}{diag}
\DeclareMathOperator{\Proj}{Proj}
\DeclareMathOperator{\Imm}{Ran}
\DeclareMathOperator{\Pf}{Pf}

\DeclareMathOperator{\Ch}{Ch}

\theoremstyle{plain}
  \newtheorem{theorem}{Theorem}[section]
  \newtheorem{lemma}[theorem]{Lemma}
  \newtheorem{proposition}[theorem]{Proposition}

\theoremstyle{definition}
  \newtheorem{definition}[theorem]{Definition}

\theoremstyle{remark}
  \newtheorem{remark}[theorem]{Remark}

\title[Splitting obstructions and $\Z_2$ invariants in TRS TIs]{Splitting obstructions and $\Z_2$ invariants in\\ time-reversal symmetric topological insulators}
\author[A.~Ferreri, D.~Monaco, G.~Peluso]{Alessandro Ferreri \and Domenico Monaco \and Gabriele Peluso}
\date{\today}

\usepackage{hyperref}
\hypersetup{colorlinks=true, linkcolor=black, citecolor=black, filecolor=black, urlcolor=black}

\begin{document}

\begin{abstract}
The Fu--Kane--Mele $\Z_2$ index characterizes two-dimensional time-reversal symmetric topological phases of matter. We shed some light on some features of this index by investigating projection-valued maps endowed with a fermionic time-reversal symmetry.

Our main contributions are threefold. First, we establish a splitting theorem, proving that any such projection-valued map admits a splitting into two projection-valued maps that are related to each other via time-reversal symmetry. Second, we provide a complete homotopy classification theorem for these maps, thereby clarifying their topological structure. Third, by means of the previous analysis, we connect the Fu--Kane--Mele index to the Chern number of one of the factors in the previously-mentioned decomposition, which in turn allows to exhibit how the $\Z_2$-valued topological obstruction to constructing a periodic and smooth Bloch frame for the projection-valued map, measured by the Fu--Kane--Mele index, can be concentrated in a single pseudo-periodic Kramers pair.
\end{abstract}

\maketitle

\tableofcontents


\section*{Introduction}

Since the experimental discovery of the quantum Hall effect in the 1980s~\cite{von1980new} and its subsequent topological interpretation~\cite{PhysRevLett.51.2167}, topological phases of matter have remained a central topic in condensed matter physics and mathematical physics. The theoretical prediction of the quantum spin Hall effect~\cite{Kane_2005}, the establishment of the “periodic table” of topological insulators and superconductors by Kitaev~\cite{Kitaev_2009}, and the first experimental confirmations~\cite{Ko_nig_2007,Ando2013} in the early 2000s have directed attention towards \textit{Topological Insulators} (TIs), materials realizing symmetry-protected topological phases.

Kitaev’s classification, derived from $K$-theoretical arguments, divides the problem into ten symmetry classes, corresponding to the so-called “tenfold way”, first formulated independently by Schnyder, Ryu, Furusaki, and Ludwig~\cite{Schnyder_2008}. This framework provides a unifying picture of topological phases characterized by different combinations of fundamental symmetries: time-reversal, particle-hole, and chiral. Among these, the class AII—comprising systems with fermionic time-reversal symmetry—has attracted particular attention due to its rich topological structure and its physical realizations in two- and three-dimensional topological insulators.

Class AII corresponds to systems endowed with an antiunitary time-reversal symmetry $\Theta$ satisfying $\Theta^2=-1$. For crystalline and insulating systems, described by a periodic Hamiltonian with a spectral gap, one associates the so-called \emph{Fermi projection} $P$, which defines the occupied-energies subspace. As reviewed below, this in turn gives rise to the \emph{Bloch bundle} $E \to \T^d$ over the Brillouin torus $\T^d$, carrying an induced antiunitary equivariant structure. Fiber bundles subject to this type of symmetry are called ``Quaternionic''-vector bundles, and have became object of independent geometric interest, beyond the specific case considered in this paper, due to their peculiar topological structure (see, e.g.,~\cite{de2015classification, bunk2019topologicalinsulatorskanemeleinvariant}).

As shown in~\cite{panati2007triviality, monaco2015symmetry}, in two-dimensional systems (to which we restrict for the rest of this Introduction) the presence of $\Theta$ guarantees the existence of a global smooth trivialization of $E$, corresponding to the vanishing of the \emph{Chern number} $\Ch(P) \in \Z$. However, such a trivialization need not respect the time-reversal symmetry itself, and the obstruction to the existence of a \emph{symmetric} trivialization carries non-trivial topological content. This obstruction is quantified by a $\Z_2$-valued invariant, first introduced in~\cite{kane2005z}. While the original formulation of this invariant was already deeply topological in nature, several alternative formulations and interpretations have since been proposed, both in the physical and in the mathematical literature, see e.g.~\cite{graf2013bulk, de2015classification, fiorenza2016z, monaco2015symmetry, FuKane2006, MooreBalents2007}. For an overview of the relation among some of the different mathematical approaches, we refer the reader to~\cite{Cornean_2017}.

Following the approach initiated in~\cite{fiorenza2016z}, the $\mathbb{Z}_2$ invariant can be interpreted as the obstruction to the existence of a globally symmetric trivialization of the Bloch bundle. In position space, the analog of a globally smooth trivialization of the Bloch bundle is the existence of exponentially localized Wannier bases. Such bases have a clear physical interpretation as representing localized electronic orbitals, and their existence is intimately related to the insulating nature of the system. The requirement that a Wannier basis is symmetric under time-reversal symmetry allows to implement the symmetry at the level of such orbitals, which can be useful when these are used, for example, to derive symmetric tight-binding models from continuum Hamiltonians~\cite{marzari2012maximally, marrazzo2024wannier}. In view of their relevance to computational solid state physics, we comment in Section~\ref{sec:parseval} on how the topological obstruction to localization of time-reversal symmetric Wannier orbitals can be avoided if one relaxes the notion of orthonormality to that of forming a \emph{Parseval frame} (compare~\cite{cornean2019parseval}).

If, in addition to $\Theta$, the system possesses a unitary involutive symmetry $S$ that is compatible with $\Theta$ ---corresponding, for example, to conservation of one component of spin--- then the topological invariant persists and admits a simple physical interpretation as the \emph{spin Chern number}~\cite{Sheng_2006,Prodan_2009}
\[
\Ch_S(P) := \frac{\Ch\big(P \tfrac{S+\Id}{2}\big) - \Ch\big(P \tfrac{S-\Id}{2}\big)}{2}.
\]
As shown by Prodan~\cite{Prodan_2009}, the spin Chern number is stable under small perturbations, remaining well defined even when the symmetry $S$ is slightly broken -- precisely, as long as the operator $PSP$ has a gap around $0$. Moreover, the same work discusses how, upon changing the spin-type symmetry $S$, the value of the spin Chern number can change only modulo $2$, making it a natural candidate for the invariant associated with the AII symmetry class.

The role of the symmetry $S$ in computing the Spin-Chern number is essentially to split the Fermi projection into two components exchanged by $\Theta$:
\begin{equation}
P = P^+ + P^-\,, \qquad P^{\pm} := \frac{1}{2}\big(P \pm P S\big),\qquad \Theta P^{+}\Theta^{-1}=P^{-}.
\end{equation}
The spin Chern number then coincides with the Chern number of either component (which necessarily agree with each other). We will refer to such a decomposition of $P$ as a \emph{symmetric splitting}. The existence of a symmetric splitting may be viewed as a weaker form of the existence of a spin--type symmetry. In the presence of such a splitting, one can still define an integer-valued invariant, which moreover enjoys the same robustness and locality properties as the Chern number. This perspective has already been partially explored in the physics literature, for instance in~\cite{Gilardoni_2022, Ba__2024}.

In the present work, we prove that, for periodic systems endowed with fermionic time-reversal symmetry it is always possible to find a symmetric splitting. We show that the topological characterization of such systems is captured by the Chern number, modulo $2$, of one component of the splitting, or equivalenly by the Spin-Chern number of $P$ with respect to a spin-type operator taylored around the splitting. The $\mathbb{Z}_2$ nature of the invariant arises from the non-uniqueness of this splitting or of this spin-type symmetry. 

Our proof of this result passes through the construction of an appropriate trivialization of the Bloch bundle, as mentioned above, which is compatible with time-reversal symmetry but possibly not with the full lattice periodicity. This construction unlocks also the possibility to rephrase the $\Z_2$ index as an invariant of the homotopy and unitary equivalence classes (and even of the Murray--von Neumann equivalence class; compare Definition~\ref{def:Equivalence relations for time-reversal symmetric projection-valued maps}) of the projection $P$, showing how the invariant is stable with respect to continuous deformations of the physical model -- that is, of its Hamiltonian -- which keep the spectral gap open, as well as with respect to unitary changes of representation which cannot affect the description of the physical properties of the system.

These results provides a new perspective on the geometric and physical meaning of the Fu–Kane–Mele index: in particular, unlike other topological formulations, it admits a clear position-space counterpart, namely the Chern marker of one of the splitting projections. This position-space counterpart of the invariant paves the way for extending several results previously obtained in the spin-conserving case to the more general symmetry class AII, such as the addition of disorder and impurities in the crystal~\cite{bellissard1994noncommutative}.

The paper is structured as follows. The first section is devoted time-reversal symmetric and periodic Hamiltonian, and how their gapped spectral eigenprojections allow to define time-reversal symmetric periodic projection-valued maps on the (Brillouin) torus. The second section revisits parallel transport in the context of continuous, rather than differentiable, projection-valued maps. The third section reviews some known results on projection-valued maps in absence of time-reversal symmetry, including the construction of continuous frames (i.e.\ trivializations) for them and how, in $d=2$, the Chern numbers obstructs the possibility to impose periodicity on the frame. The fourth section introduces the $\Z_2$-valued topological invariant for two-dimensional time-reversal symmetric projection-valued maps, which has a similar interpretation as a topological obstruction. The main results of the paper, regarding this $\Z_2$ invariant, are collected in the fifth section, and present its relation to symmetric splittings, the Spin-Chern number, and the completeness of the invariant as a marker for the different equivalences discussed above. Finally, the appendix presents a series of classical results in basic algebraic topology that are used throughout the paper.

\subsubsection*{Acknowledgements.} A.~F. and G.~P.\  would like to thank the organizers of the {\it Quantum Mathematics 
@ PoliMI Intensive Period} for the kind hospitality. The Authors also thank Marco Valerio D'Agostino and Gianluca Panati for valuable discussions on topics related to the paper.  

D.~M. and G.~P.\ gratefully acknowledge financial support from Sapienza Universit\`{a} di Roma within Progetto di Ricerca di Ateneo 2023 and 2024, as well as from MUR--Italian Ministry of University and Research and Next Generation EU within PRIN 2022AKRC5P ``Interacting Quantum Systems: Topological Phenomena and Effective Theories''.  The Authors gratefully acknowledge financial support from PNRR MUR under Project No. PE0000023-NQSTI.

This work has been carried out under the auspices of the GNFM-INdAM (Gruppo Nazionale per la Fisica Matematica --- Istituto Nazionale di Alta Matematica).

\section{From (symmetric) periodic Hamiltonians to (symmetric) projection-valued maps} 

In this section, we outline the standard procedure that allows one to associate a projection-valued map to a periodic Hamiltonian, following the construction reviewed in~\cite{panati2007triviality, panati2013bloch} to which we refer the reader for further details. We consider a quantum particle moving in a condensed matter system, whose configuration space $X \subset \R^d$ exhibits a periodic structure with respect to a lattice $\Lambda \simeq \Z^d$, commonly referred to as the \emph{Bravais lattice}. Explicitly, this periodicity implies $X = \lambda + X$ for every $\lambda \in \Lambda$. Two types of models are typically used in this context: \emph{continuum models}, in which $X=\R^d$, and \emph{tight-binding models}, where $X$ is replaced by the discrete set of atomic positions and the quotient $X/\Lambda$ is finite. When~$n$ additional internal degrees of freedom, such as spin, are present, the configuration space is naturally extended to $X \times \{1,\ldots,N\}$, and the corresponding Hamiltonian $H$ describing the system acts on $L^2(X)\otimes \C^N$. Furthermore, the system is assumed to be endowed with a \emph{fermionic time-reversal symmetry}, i.e., an anti-unitary operator $\TRS$ acting on $L^2(X)\otimes \C^N$ such that $\TRS^2 = -\Id$, commuting with the Hamiltonian $H$. The discussion therefore puts on the same footing both continuum models ($X=\R^d$), described by Hamiltonians with spin-orbit coupling of Rashba type~\cite{rashba2005spin}, and discrete, tight-binding models, like the paradigmatic Kane--Mele Hamiltonian~\cite{Kane_2005, kane2005z}. For the sake of presentation, we'll start from setting up the mathematical framework in continuum models: compare Remark~{rmk:discrete} for the minor changes that take place in the discrete setting.

The periodicity of the system is implemented at the operator level as a commutation relation between $H$ and lattice translations,
\[
H T_\lambda = T_\lambda H,
\]
where $(T_\lambda \phi)(x) = \phi(x-\lambda) \in \C^N,$ for all $\lambda \in \Lambda$. Correspondingly, the fermionic time-reversal symmetry operator $\Theta$ is required to be itself translation-invariant, i.e., to commute with the lattice translations $T_\lambda$.  

A powerful tool to study the spectral properties of such periodic Hamiltonians is the \emph{Bloch-Floquet-Zak transform}~\cite{kuchment2016overview, panati2013bloch}, which allows one to decompose the Hilbert space into fibers labeled by the quasi-momentum $k$. For any sufficiently rapidly decaying $\phi \in L^2(\R^d) \otimes \C^N$, the transform is defined as the following function depending on $k \in \R^d$ and $y \in \R^d$:
\[
(\UZ \phi)_k(y) = \sum_{\lambda \in \Lambda} e^{-ik\cdot (\lambda - y)} (T_\lambda \phi)(y).
\]

Let us denote by $\Lambda^* = \{\lambda^* \in \R^d \mid \lambda^* \cdot \lambda \in 2\pi \Z, \forall \lambda \in \Lambda \}$ the dual lattice, and by $\mathbb{B}$ a connected periodicity cell, the \emph{Brillouin zone}, for $\Lambda^*$. With this choice, $(\UZ \phi)_k(y)$ naturally belongs to a fiber Hilbert space with constant fibers:
\[
(\UZ \phi)_k(y) \in \left\{ f \in L^2_{\rm loc}(\R^d, dk, \Hi_f) \ \middle| \ f(k+\lambda^*) = \tau_{\lambda^*} f(k) \right\} =: \Hi_\tau \simeq \int_{\mathbb{B}}^\oplus dk \, \Hi_f,
\]
where $\Hi_f$ is the space of $\C^N$-valued, $\Lambda$-periodic, locally square-integrable functions on $\R^d$, and $\tau_{\lambda^*}:\Hi_f \to \Hi_f$ acts as $\tau_{\lambda^*}(g)(y) = e^{i \lambda^* \cdot y} g(y)$. By also choosing a periodicity cell for $\Lambda$, the \emph{Wigner-Seitz cell} $\mathbb{W}$, one can further identify $\Hi_f$ with $L^2_{\rm per}(\mathbb{W}) \otimes \C^N$, the space of $\C^N$-valued square-integrable functions on $\mathbb{W}$ with $\Lambda$-periodic boundary conditions. A classical result, which can be traced back to~\cite{bloch1929quantenmechanik, zak1964magnetic} (compare~\cite{reed1978iv}), reads:

\begin{proposition}\label{pro:BFZ-unitary}
The operator $\UZ$ defined above extends uniquely to a unitary operator
\[
\UZ : L^2(\R^d) \otimes \C^N \longrightarrow \int_{\mathbb{B}}^\oplus dk \, L^2_{\rm per}(\mathbb{W}) \otimes \C^N.
\]
\end{proposition}

When $H$ acting in $L^2(\R^d) \otimes \C^N$ commutes with the translation operators $T_\lambda$, $\lambda \in \Lambda$, the transformed operator $\UZ \, H \, \UZ^{-1}$ is \emph{decomposable}: there exists a family of self-adjoint \emph{fiber operators} $\{H(k)\}_{k \in \R^d}$ such that
\[
\big[ \UZ\, H \,\UZ^{-1} f \big]_k = H(k) f_k, \quad \forall f \in \Hi, \ k \in \R^d,
\]
and these fibers satisfy the $\tau$-covariance condition
\begin{equation}\label{eq:tau-covariance_Hamiltonian}
H(k+\lambda^*) = \tau_{\lambda^*} H(k) \tau_{\lambda^*}^{-1}, \quad \forall k \in \R^d, \ \lambda^* \in \Lambda^*.
\end{equation}

The same procedure can be applied to $\TRS$, since it commutes with the translation operators. Consequently, there exists a family of anti-unitary operators $T(k)$ acting on $\Hf$ such that
\[
\left[\UZ \, \TRS \,\UZ^{-1} (f)\right]_k = T(k) f_{-k}, \quad \forall f \in \Hi_\tau, \ k \in \R^d.
\]
The appearance of the minus sign in front of $k$ is a direct consequence of the anti-linearity of $\TRS$, and reflects also on the $\tau$-covariance condition:
\[
T(k+\lambda^*) = \tau_{\lambda^*} T(k) \tau_{\lambda^*}, \quad \forall k \in \R^d, \ \lambda^* \in \Lambda^*.
\]

In most physical models, $\TRS$ acts only on the second factor of the tensor product $L^2(X)\otimes \C^N$, which implies that the fibers $T(k)$ are independent of $k$. Henceforth, we will assume $T(k) \equiv T$. Although it is possible to generalize the analysis to a general $\TRS$, this extension lies beyond the scope of the present work. Straightforward computations then show that the commutation relation $TH = HT$ at the level of fibers reads
\begin{equation}\label{eq:TRS_fiber_Hamiltonians}
TH(k) = H(-k) T, \quad \forall k \in \R^d.
\end{equation}

Let us return to the fiber Hamiltonians. In most physically relevant models, the mapping $k \mapsto H(k)$ constitutes an \emph{analytic family of operators} in $\Hf$ in the sense of Kato~\cite{kato2013perturbation}, meaning in particular that the resolvent operators $(H(k)-z)^{-1}$ are bounded operators depend analytically on the momentum $k$; moreover, the spectrum of the full Hamiltonian satisfies $\sigma(H) = \bigcup_{k \in \R^d} \sigma
(H(k))$~\cite{reed1978iv}. 
Typically, $\sigma(H)$ consists of disjoint closed intervals, referred to as \emph{Bloch bands}, which can be used, as shown by the following proposition, to construct projection-valued maps, which will be the main object of study in the following sections.

\begin{proposition}[Riesz formula]\label{pro:Riesz-formula}
Let $H(k)$ be a family of fiber Hamiltonians, and let $\Omega$ be a set of Bloch bands that is spectrally separated from the rest of the spectrum. Let $\gamma$ be a smooth, closed, positively oriented contour contained in the resolvent set of $H(k)$ for all $k$, enclosing $\Omega$ and no other part of the spectrum. Then
\[
P_\Omega(k) = \frac{i}{2\pi} \int_\gamma \left(H(k) - z \Id \right)^{-1} \, dz
\]
defines a family of projections on $\Hf$ satisfying $P_\Omega(k) = \chi_\Omega(H(k))$, where $\chi_\Omega$ denotes the spectral projection onto $\Omega$. Moreover, the map $k \mapsto P_\Omega(k)$ inherits regularity in $k$, $\tau$-covariance, and (if present) time–reversal symmetry from the map $k \mapsto (H(k)-z)^{-1}$.

\end{proposition}
\begin{proof}
Observe that if $f(w) = \frac{i}{2\pi} \int_\gamma \frac{1}{w-z} dz$, then $P_\Omega(k) = f(H(k))$. The integral computes the winding number of $\gamma$ around $w$, which is 1 if $w$ lies inside $\gamma$ and 0 otherwise. By assumption, the spectrum of $H(k)$ inside $\gamma$ is contained in $\Omega$. Therefore,
\[
P_\Omega(k) = f(H(k)) = \chi_\Omega(H(k)).
\]

The regularity statement follows immediately, since derivatives can be brought inside the integral.
\end{proof}

An immediate corollary is that the rank of $P_\Omega(k)$ is constant ---say, equal to $m \in \N$---, since $k \mapsto \tr(P_\Omega(k)) = \dim \Imm(P_\Omega(k))$ defines a continuous map into a discrete set. Henceforth, we denote by $\Proj_m(\Hf)$ the space of rank-$m$ on the fiber Hilbert space $\Hf$.

Let us spell out $\tau$-covariance and the time-reversal symmetry constraint inherited from Equations~\eqref{eq:tau-covariance_Hamiltonian} and~\eqref{eq:TRS_fiber_Hamiltonians}:
\[
P_\Omega(k+\lambda^*) = \tau_{\lambda^*} P_\Omega(k) \tau_{\lambda^*}^{-1}, \quad
T P_\Omega(k) = P_\Omega(-k) T, \quad \forall k \in \R^d, \ \lambda^* \in \Lambda^*.
\]
Although the $\tau$-covariance of this family of projections is essential for defining the topological invariants we aim to study, the associated pseudo-periodicity can be cumbersome to handle. To address this, we employ the following construction, first introduced in~\cite{cornean2016construction}.
We begin by considering a lattice basis $\{\lambda_1^*, \lambda_2^*, \dots, \lambda_d^*\}$ for $\Lambda^*$ and the corresponding unitary operators $\tau_{\lambda_j^*}(\phi)(y) = e^{i \lambda_j^* \cdot y} \phi(y)$ acting on $\Hi$.  
Using the spectral theorem for unitary operators, we find $d$ self-adjoint operators $L_1, \dots, L_d$ such that $\tau_{\lambda_j^*} = e^{i L_j}$. Since the $\tau_{\lambda_j^*}$'s commute with each other, they can be chosen to satisfy $[L_a, L_b] = 0$ for all $a,b \in \{ 1, \dots, d\}$.  

The $\tau$-covariance of the fibers of $\TRS$ implies
\[
T(k+\lambda_j^*) = \tau_{\lambda_j^*} T(k) \tau_{\lambda_j^*}.
\]  
If the fibers are constant, this reduces to $T e^{i L_j} = e^{-i L_j} T$; by functional calculus one can choose $L_j$ such that $T L_j = -L_j T$, so that
\[
T e^{i L_j k_j} = e^{-i L_j k_j} T, \quad \forall k_j \in \R.
\]

We can now define a unitary-valued map $U:\R^d \to \U(\Hf)$ as
\[
U(k_1, \dots, k_d) = \prod_{j=1}^d e^{i L_j k_j}.
\]  
By construction, it satisfies
\[
T U(k_1, \dots, k_d) = U(-k_1, \dots, -k_d) T.
\]  
If $(k_1, \dots, k_d) = k$ are the coordinates of $k \in \R^d$ in the chosen lattice basis, we define
\[
\Tilde{P}_\Omega(k) = U(k_1, \dots, k_d) P_\Omega(k) U(k_1, \dots, k_d)^{-1} \equiv U(k) P_\Omega(k) U(k)^{-1}.
\]  
It is immediate to check that this projection-valued map is now $\Lambda^*$-periodic in $k$ while still satisfying the symmetry constraint
\[
\Tilde{P}_\Omega(k) T = T \Tilde{P}_\Omega(-k).
\]  

The choice of $L_1, \dots, L_d$ is not unique: if another unitary-valued map $U':\R^d \to \U(\Hf)$ has the same properties as the $U$ constructed above, then the corresponding projections $\Tilde{P}_\Omega$ and $\Tilde{P}'_\Omega$ are unitarily equivalent via the periodic unitary-valued map $U'(k) U(k)^{-1}$:
\[
\Tilde{P}'_\Omega(k) = U'(k) U(k)^{-1} \Tilde{P}_\Omega(k) \left[ U'(k) U(k)^{-1} \right]^{-1},
\]  
with the conjugation map satisfying the time-reversal symmetry constraint
\[
T U'(k) U(k)^{-1} = U'(-k) U(-k)^{-1} T, \quad \forall k \in \R^d.
\]

With this construction, we obtain a regular projection-valued map $P:\R^d \to \Proj_m(\Hf)$ that is $\Lambda^*$-periodic in $k$. We may reinterpret it as a projection-valued map $P:\T^d \to \Proj_m(\Hf)$ since $\R^d/\Lambda^* \simeq \T^d$, which furthermore satisfies the symmetry constraint
\[
T P(k) = P(-k) T, \quad \forall k \in \T^d.
\]
A family of projections $P$ satisfying the above assumptions will be the central object of study in this article. 

\begin{remark} \label{rmk:discrete}
Before giving a formal definition, we briefly consider again the discrete setting, where the Bloch-Floquet transform and the procedure to reduce $\tau$-covariance to $\Lambda^*$-periodicity can be summarized in a single step. As we will see, we obtain again a projection-valued map that is periodic and satisfies a symmetry constraint analogous to the one above.

Let $X \subset \R^d$ be a discrete configuration space which is invariant under shifts in a lattice $\Lambda \simeq \Z^d$, and suppose $X/\Lambda$ is finite. We can then select a finite set of points $\mathbb{W} = \{x_1, \dots, x_l\} \subset X$ such that $X = \mathbb{W} + \Lambda$. Let $H$ be an Hamiltonian acting in $L^2(X) \otimes \C^n$, which we rewrite as
\[
L^2(X) \otimes \C^N \simeq L^2(\mathbb{W} \times \Lambda) \otimes \C^N \simeq \ell^2(\Lambda) \otimes \C^{l \cdot N} \simeq \ell^2(\Z^d) \otimes \C^{l \cdot n}.
\]  
We denote by $\mathcal{W}$ the corresponding isomorphism, which depends on the choice of $\mathbb{W}$. Applying the Fourier transform
\[
\F : \ell^2(\Z^d) \to L^2(\T^d, dk, \C), \quad (\F f)(k) = \sum_{\lambda \in \Z^d} \frac{1}{(2\pi)^{d/2}} e^{i k \cdot \lambda} f(\lambda),
\]  
to the first factor, the operator
\[
(\F \otimes \Id) \mathcal{W} H \mathcal{W}^{-1} (\F \otimes \Id)^{-1}, \quad \text{acting on } L^2(\T^d, dk, \C^{l \cdot N}) \simeq \int_{\T^d}^\oplus dk \, \C^{l \cdot N},
\]  
is decomposable into a family of self-adjoint fibers $\{H(k)\}_{k \in \T^d}$, in the sense specified above. Here, periodicity in $k$ holds directly, which is why we write $k \in \T^d$. The same steps as in the continuum case yield a projection-valued map $P:\T^d \to \Proj_m(\C^{l\cdot n})$ satisfying
\[
T P(k) = P(-k) T, \quad \forall k \in \T^d,
\]  
assuming constant fibers of $\TRS$.

It is important to note that the choice of discrete Wigner-Seitz cell $\mathbb{W} = \{x_1, \dots, x_l\}$ is not unique, and this choice could in principle affect the topological properties of the model. If a different cell $\mathbb{W}' = \{x_1', \dots, x_l'\}$ is chosen, one defines another isomorphism $\mathcal{W}'$ and obtains a new fiber decomposition
\[
\big[ (\F \otimes \Id) \mathcal{W}' H (\mathcal{W}')^{-1} (\F \otimes \Id)^{-1} (f) \big]_k = H'(k) f_k, \quad \forall f \in L^2(\T^d, dk, \C^{l \cdot N}), \ k \in \T^d.
\]  
The two decompositions are related by a fibered unitary equivalence
\[
(\F \otimes \Id) \mathcal{W}' \mathcal{W}^{-1} (\F \otimes \Id)^{-1} = \int_{\T^d}^\oplus dk \, U(k),
\]  
where $U(k)$ acts on $\C^{l \cdot N}$, so that
\[
U(k) H(k) U(k)^{-1} = H'(k),
\]  
and likewise for the associated projection-valued maps $P_\Omega(k)$ and $P_\Omega'(k)$. To preserve the symmetric structure, one must choose $\mathbb{W}$ such that
\[
U(k) T = T U(-k), \quad \forall k \in \T^d.
\]
\end{remark}

In both continuum and discrete settings, we thus obtain the central object of study in this paper.

\begin{definition}[Time-reversal symmetric projection-valued map]\label{def:TRSPVM}
Let $d \in \N$ (the physical dimension of the model), $m \in \N$ (the rank of the projections), $\Hi$ a Hilbert space, and $T$ an anti-unitary operator on $\Hi$ with $T^2 = -\Id$, $m \in \N$. A continuous, periodic projection-valued map $P:\T^d \to \Proj_m(\Hi)$ is \emph{time-reversal symmetric} if
\begin{equation} \label{eq:TRSPVM}
T P(k) = P(-k) T, \quad \forall k \in \T^d.
\end{equation}
\end{definition}

Notice in particular that, on points $k_\star$ fixed by the involution $k \mapsto -k$ on the torus $\T^d$, the antiunitary operator $T$ induces the structure of an $m$-dimensional quaternionic Hilbert space on the range of the projection $P(k_\star)$, as $T P(k_\star) = P(k_\star) T$. Applying Lemma \ref{lem:anti-unitary_quaternionic_structure} from Appendix \ref{section: Tools from algebraic geometry} yields the result that the dimension of $\Imm(P(k_\star))$ must be even. Since the rank is constant in $k$, we have that $m \in 2\Z$. So from now on we impose $m=2n$.

\begin{definition}[Symmetric periodic (Bloch) frame]
A \emph{frame} for a projection-valued map $\{P(k)\}_{k \in \R^d}$ is a collection of orthonormal vectors $\{v_j(k)\}_{j\in\{1,\cdots, m\}}$ that are continuous in $k$ and span $\Imm(P(k))$ orthonormally for all $k \in \R^d$.

If the projection-valued map $\{P(k)\}_{k \in \T^d}$ is $\Lambda$-periodic, the frame is called \emph{periodic} if the vectors $v_j(k)$ depend $\Lambda$-periodically on $k$.

If the projection-valued map is time-reversal symmetric, and therefore $m=2n$, the frame is called \emph{symmetric} if
\[ T v_j(k) = v_{j+n}(-k) \quad \text{for all } 1 \le j \le n. \]
\end{definition}

Notice how the symmetry constraint couples vectors in a frame into \emph{Kramers pairs}, and it is not possible to relate a vector to itself by imposing $Tv(k)=v(-k)$. In fact, this would lead to $Tv(0)=v(0)$ and consequently to $-v(0)=T^2v(0) =v(0)$ which is absurd. This simpler symmetry relation can instead be imposed when $T^2=\Id$, as discussed in~\cite{cornean2016construction} and~\cite{fiorenza2016z}.

The existence of a symmetric and periodic frame is not guaranteed in general for a time-reversal symmetric projection-valued map $P$, and if $d=2$ the obstruction to constructing one, as explained in Section \ref{section: The symmetric matching matrix}, is measured by the $\mathbb{Z}_2$ invariant associated with $P$. In Theorem \ref{thm:equivalence_theorem} we will show that, instead, it is always possible to construct a \emph{pseudo-periodic} symmetric frame in the sense of the following definition.

\begin{definition}[Pseudo–periodic symmetric frame]\label{def:pseudo-periodic symmetric frame}
    Let $P$ be a rank-$2n$ time-reversal symmetric projection-valued map in the sense of Definition \ref{def:TRSPVM}. A frame $\{v_j\}_{1 \le j\leq 2n}$ is called \emph{pseudo-periodic} if the vectors define continuous maps from $[-\pi,\pi]\times \T^1$ to $\Hi$ that satisfy 
    \begin{enumerate}
        \item periodicity on $2n-2$ elements, i.e. 
        $$
            v_j(\pi,k_2)=v_j(-\pi,k_2)\quad \forall k_2\in\T^1,\qquad  \forall j\notin\{1, n+1\},
        $$
        \item and periodicity up to a phase on $2$ elements, i.e. for some $h \in \Z$
        $$
            v_1(\pi,k_2)=e^{ih k_2}v_1(-\pi,k_2)\quad\text{and}\quad
            v_{n+1}(\pi,k_2)=e^{-ih k_2}v_{n+1}(-\pi,k_2)\quad \forall k_2\in\T^1.$$
    \end{enumerate}
 
 A pseudo-periodic frame is called \emph{symmetric} if the vectors further satisfy
 \begin{enumerate}[resume]
 \item time-reversal symmetry, i.e.
        $$T v_i(t,k_2)=v_{n+i}(-t,-k_2), \quad \forall t\in[-\pi,\pi],\ k_2\in\T^1.$$
 \end{enumerate}
\end{definition}

In Theorem \ref{thm:equivalence_theorem}, the existence of a pseudo-periodic symmetric frame will be related to the existence of a decomposition of the projection-valued map $P$ that preserves the symmetry in an appropriate sense, namely

\begin{definition}[Symmetric splitting]\label{def:symmetric_splitting}
     Let $P$ be a rank-$2n$ time-reversal symmetric projection-valued map in the sense of Definition \ref{def:TRSPVM}. A \emph{symmetric splitting} of $P$ is a pair of continuous projection-valued maps $P^\pm:\T^d \to \Proj_{n}(\Hi) $ such that 
\begin{equation*}
    P^+(k)P^-(k)\equiv 0, \quad P(k)=P^+(k)+P^-(k) \quad \mbox{and} \quad TP^+(k)=P^-(-k)T \quad \forall k \in \T^d.
\end{equation*}
\end{definition}

The usual equivalence relations for projection-valued maps, which typically arise in the study of $K$-theory~\cite{rordam2000introduction}, can be adapted to the time-reversal symmetric setting. It is with respect to these adapted equivalence relations that we will study the completeness of the invariant defined in Section \ref{section: The symmetric matching matrix}.

\begin{definition}[Equivalence relations for time-reversal symmetric projection-valued maps]\label{def:Equivalence relations for time-reversal symmetric projection-valued maps}
    Let $P_1, P_2$ be two time-reversal symmetric projection-valued maps in the sense of Definition~\ref{def:TRSPVM}. We say that $P_1$ and $P_2$ are:
\begin{enumerate}
    \item \emph{Murray–von Neumann equivalent} if there exists a continuous map $V:\T^d \to \mathcal{B}(\Hi)$ with values in bounded operators such that
    \begin{equation*} \label{eq:TRS_Murray_equivalence}
    T V(k) = V(-k) T,\quad 
    V(k)^*V(k) = P_0(k),\quad 
    V(k)V(k)^* = P_1(k),\quad 
    \forall k \in \T^d
    \end{equation*}
    (namely, $V(k)$ is a partial isometry between the ranges of $P_0(k)$ and $P_1(k)$ compatible with time-reversal symmetry);
    \item \emph{unitarily equivalent} if there exists a continuous unitary-valued map $U:\T^d \to \U(\Hi)$ such that
    \begin{equation*} \label{eq:TRS_unitary_equivalence}
    T U(k) = U(-k) T,\quad 
    P_1(k) = U(k) P_0(k) U(k)^{-1},\quad 
    \forall k \in \T^d;
    \end{equation*}

    \item \emph{homotopic} if there exists a continuous projection-valued map $P:[0,1] \times \T^d \to \Proj_m(\Hi)$ such that
    \begin{equation*}\label{eq:TRS_homotopy}
    P(0,k) = P_0(k),\quad 
    P(1,k) = P_1(k),\quad 
    T P(t,k) = P(t,-k) T,\quad 
    \forall t\in[0,1],\ k\in\T^d.
    \end{equation*}
\end{enumerate}
\end{definition}

\begin{remark}[Role of Murray--von Neumann, unitary and homotopy equivalences]\label{rmk:Varie equivalenze}
As mentioned above, the three notions introduced in the previous Definition are natural equivalence relations that can be defined for projections in any $C^*$-algebra. In the present setting, the relevant $C^*$-algebra is the one of maps $A(k)$ from the torus $\mathbb{T}^2$ to $\mathcal{B}(\mathcal{H})$ satisfying a symmetry constraint of the form 
$$
TA(-k)=A(k)T, \quad \forall k\in \T^2\,.
$$

Under suitable assumptions (for instance, when $\mathcal{H}$ is infinite-dimensional) these equivalence notions coincide. In general, however, unitary equivalence corresponds to  Murray--von Neumann equivalence of both the projections and their orthogonal complements, while homotopy equivalence corresponds to unitary equivalence implemented by unitaries homotopic to the identity -- unless the unitary group of the $C^*$-algebra is connected (which is indirectly proven in Lemma \ref{lem:infinite_dim_Class_AII_frame_d=2}), homotopy is strictly finer than the previous two. For a detailed discussion of the hierarchy among these three equivalence relations, valid for any $C^*$-algebra, we refer the reader to~\cite[Section 2.2]{rordam2000introduction}.

From a differential-geometric viewpoint, Murray--von Neumann equivalences correspond precisely to isomorphisms between the vector bundles associated with the projection-valued maps: physically, they would correspond to changes of representation on the Hilbert space of the quantum system which affect only states in the occupied energy levels, below the spectral gap of the Hamiltonian.
Instead, unitary equivalences are physically associated, for example, with the different choices of dimerization (that is, of the Wigner--Seitz cell) in tight-binding models, or with the selection of a unitary-valued map that removes the $\tau$-covariance in continuum models. 
As for homotopies, they correspond to continuous deformations of physical parameters in the original Hamiltonian that do not close the spectral gaps. Under such deformations, the topological properties of $P$ are preserved.  
Since the physical interpretations of the last two mathematical notions differ, it is natural to treat them separately. Moreover, when the two conditions coincide mathematically, they establish an equivalence between distinct physical aspects of the system.
\end{remark}

\begin{remark}[Continuous vs. analytic frames]\label{rmk:continuous_vs_analytic}

Since the family of projections $\{P(k)\}_{k \in \R^d}$ arising in physical models is usually analytic in $k$, one may ask whether (real-)analytic frames can be constructed. This enhanced regularity on the frames would correspond, in position space, i.e.\ via the inverse Bloch--Floquet(--Zak) transform, to exponential localization of the corresponding Wannier functions, as mentioned in the Introduction. Following the argument of~\cite[Lemma 2.3]{cornean2016construction}, one can show that whenever a continuous frame exists, it can be modified to a real-analytic one while preserving the time-reversal symmetry.
\end{remark}

\section{Continuous parallel transport}

From here on out, we focus on $2$-dimensional systems, and set $d=2$ once and for all. 

One typical approach to the construction of a frame for a given projection-valued map $\{P(k)\}_{k \in \R^2}$ is based on the following observation: if one can unitarily intertwine two projections in the family, i.e.\ if $P(k') = U \, P(k) \, U^{-1}$ for some $k, k' \in \R^2$, then an orthonormal basis for $\Imm P(k)$ can be ``transported'', via the unitary $U$, to yield a basis for $\Imm P(k')$. Therefore, in this Section we investigate the problem of constructing a unitary-valued map $U:\T^2 \to \U(\Hi)$ associated with a given projection-valued map $P:\T^2 \to \Proj_n(\Hi)$, such that
\begin{equation}\label{eq: parallel transport}
    P(k) = U(k)\, P(0)\, U(k)^{-1}.
\end{equation}
If $P$ is further time-reversal symmetric, we will also require that
\begin{equation}\label{eq: parallel transport symmetrico}
    TU(k) = U(-k)T, \quad \forall k\in \T^2,
\end{equation}
so that the time-reversal symmetry constraint can be preserved also at the level of symmetries.

As we shall see, it is in general impossible to find such a map $U$ that is continuous and periodic in both variables; periodicity can be achieved only in one of them, yielding a continuous map defined on $
[-\pi,\pi]\,\times \T^1  $
which satisfies \eqref{eq: parallel transport} and, in the presence of time-reversal symmetry, also \eqref{eq: parallel transport symmetrico}.

A standard approach to this problem relies on \emph{parallel transport}, which is widely used in the literature on symmetric projection-valued maps (see, e.g.,~\cite{cornean2016construction, cances2017robust, Cornean_2017, cornean2019parseval, gontier2019numerical, gontier2022symmetric, monaco2023az}). This method requires at least $C^1$ regularity of the map $k \mapsto P(k)$. Here, also in view of Remark~\ref{rmk:continuous_vs_analytic}, we present a method based on the \emph{Kato–Nagy intertwiner}, which applies already in the merely continuous setting and thus provides a genuine generalization of parallel transport to lower regularity, in the context of continuous projection-valued maps on the torus $\T^d$. The whole construction is based on the following Theorem, a proof of which can be found in~\cite{kato2013perturbation}:

\begin{theorem}[Kato–Nagy unitary equivalence]\label{thm:Kato-Nagy,d=0} 
If $P,Q \in \Proj_r (\Hi)$ are two projections with $\| P-Q \|< 1$, then 
\begin{equation}\label{eq:Kato-Nagy_unitary_equivalence} 
U_{P\to Q}= \left[ QP + (\Id-Q) (\Id-P)\right]\left[\Id - (P-Q)^2\right]^{-1/2} 
\end{equation}
is a unitary operator such that $Q=U_{P\to Q}PU_{P\to Q}^{-1}$.
\end{theorem}

The next result shows the existence of a periodic and time-reversal symmetric intertwiner for $1$-dimensional periodic restrictions of the projection-valued map.

\begin{proposition}
\label{pro:Kato-Nagy,d=1} 
If $P:\T^1 \to \Proj_m(\Hi)$ is a continuous projection-valued map, then there exists a continuous unitary-valued map $U: \T^1 \to \U(\Hi)$ such that
\[
P(k) = U(k) P(0) U(k)^{-1}.
\]
If $P$ is time-reversal symmetric, it is possible to choose $U$ so that $T U(k)=U(-k)T$.
\end{proposition}

\begin{proof}
Divide $[0,\pi]$ into a finite number of intervals $I_j=[t_j,t_{j+1}]$ with $t_0=0$, $t_J=\pi$, such that $\|P(t_j) -P(t) \|<1$ for all $t\in[t_j,t_{j+1}]$. For $t\in I_j$, define $U_{t_j}^t=U_{P(t_j) \to P(t)}$ according to Theorem \ref{thm:Kato-Nagy,d=0}.  
A unitary equivalence between $P(0)$ and $P(t)$, $t \in [0,\pi]$ is obtained via
\[
\Tilde{U}(t) := U_{t_j}^t U_{t_{j-1}}^{t_j} \cdots U_{t_0}^{t_1}.
\]
If $P$ is time-reversal symmetric, extend $\Tilde{U}$ to $[-\pi,0]$ via $\Tilde{U}(t)=T \Tilde{U}(-t)T$; otherwise, replicate the same procedure for $t\in[-\pi,0]$. In both cases, $\Tilde{U}:[-\pi,\pi] \to \U(\Hi)$ satisfies $P(t) = \Tilde{U}(t)P(0)\Tilde{U}(t)^{-1}$ for all $t \in [-\pi,\pi]$.  

This $\Tilde{U}$ is in general not periodic in $t$, so consider $\Tilde{U}(-\pi)^{-1}\Tilde{U}(\pi)$, which commutes with $P(0)$. If $P$ is symmetric, it also satisfies $T\left[\Tilde{U}(-\pi)^{-1}\Tilde{U}(\pi)\right] = \left[\Tilde{U}(-\pi)^{-1}\Tilde{U}(\pi)\right]^{-1} T$. By the spectral theorem, there exists a self-adjoint operator $L$ commuting with $P(0)$ such that $\Tilde{U}(-\pi)^{-1}\Tilde{U}(\pi)=e^{iL}$. If $P$ is symmetric, $L$ can also be chosen to commute with $T$. Finally, define
\[
U(t) = \Tilde{U}(t)e^{-itL/2\pi}.
\]
This construction ensures that $U(\pi) = U(-\pi)$, and therefore $U$ can be continuously extended to the whole $\R$ in a periodic way, maintaining the intertwining condition between $P(0)$ and $P(t)$, and respecting the symmetry constraint when $P$ does.
\end{proof}

As mentioned above, the latter construction fails to produce a fully-periodic unitary intertwiner, if one moves to $d=2$.

\begin{proposition}
\label{pro:Kato-Nagy,d=2}
If $P:\T^2 \to \Proj_n(\Hi)$ is a continuous projection-valued map, then there exists a continuous unitary-valued map $U:[-\pi,\pi] \times \T^1\to \U(\Hi)$, periodic in the second variable, such that
\[
P(t,k_2) = U(t,k_2)P(0,0) U(t,k_2)^{-1} \quad \forall\; (t,k_2) \in [-\pi,\pi] \times \T^1 .
\]
If $P$ is time-reversal symmetric, $U$ can be chosen such that $TU(t,k_2) = U(-t,-k_2)T$ for all $(t,k_2)\in[-\pi,\pi]\times \T^1$.
\end{proposition}

\begin{proof}
Divide $[0,\pi]$ into a finite number of intervals $ [t_j,t_{j+1}]$ with $t_0=0$, $t_J = 2\pi$, such that $\max_{k_2 \in \T^1} \| P(t_j,k_2) - P(t,k_2)\| < 1/2$ for all $(t,k_2)\in [t_j, t_{j+1}]\times \T^1$. Define $U_{t_j}^t(k_2)=U_{P(t_j,k_2) \to P(t,k_2)}$ via Theorem \ref{thm:Kato-Nagy,d=0}.  

To define the unitary equivalence between $P(0,0)$ and $P(t,k_2)$, first apply Proposition \ref{pro:Kato-Nagy,d=1} to the 1-d projection-valued map $P(0,k_2)$ to construct a periodic (and possibly symmetric) unitary-valued map $V \colon \T^1 \to \U(\Hi)$ with $P(0,k_2) =V(k_2) P(0,0)V(k_2)^{-1}$. Then, for $t\in [t_j,t_{j+1}]$, set
\[
U(t,k_2) := U_{t_j}^t (k_2) U_{t_{j-1}}^{t_j}(k_2) \cdots U_{t_0}^{t_1}(k_2) V(k_2).
\]
For $t\in[-\pi,0]$, either impose $U(t,k_2)=TU(-t,-k_2)T$ if $P$ is symmetric, or replicate the previous construction otherwise. In both cases, the claim follows.
\end{proof}

\section{Review: frames in absence of time-reversal symmetry}

In this Section, we momentarily forgo the constraint of time-reversal symmetry, and review how the construction of continuous and periodic frames for 2-dimensional periodic projection-valued maps gives rise to an integer valued topological obstruction, namely the Chern number of the family of projections. We follow the method presented e.g. in~\cite{cornean2019parseval, monaco2023topology}: we refer the reader there and to the references therein for a more detailed account. The present presentation has the advantage to fully remain in the continuous regularity class, and to set up notions and notation for the generalization to the time-reversal symmetric setting.

The aim of this Section is thus to introduce two constructions that will play a central role throughout the text, related to a rank-$m$ projection-valued map $\{P(k)\}_{k \in \T^2}$. The first provides an equivalent formulation of the Chern number in terms of the \emph{matching matrices} defined from the intertwining unitaries constructed in Proposition~\ref{pro:Kato-Nagy,d=2}. The second provides a way to build a continuous frame $\{v_j(t,k_2)\}_{1 \le j \le m}$ for $P$, $(t,k_2) \in [-\pi,\pi] \times \T^1$, that is ``as periodic as possible'' -- precisely, \emph{pseudo-periodic}, in a sense akin to that introduced in Definition~\ref{def:pseudo-periodic symmetric frame}.

\begin{definition}[Pseudo–periodic frame]\label{def:pseudo-periodic frame}
    Let $P$ be a rank-$m$ projection-valued map in the sense of Definition \ref{def:TRSPVM}. A frame $\{v_j\}_{1 \le j\leq m}$ is called \emph{pseudo-periodic} if the vectors define continuous maps from $[-\pi,\pi]\times \T^1$ to $\Hi$ that satisfy 
    \begin{enumerate}
        \item periodicity on $m-1$ elements, i.e. 
        $$
            v_j(\pi,k_2)=v_j(-\pi,k_2)\quad \forall k_2\in\T^1,\qquad  \forall j \ne 1;
        $$
        \item periodicity up to a phase on $1$ element, , i.e. for some $h \in \Z$
        $$
            v_1(\pi,k_2)=e^{i h k_2}v_1(-\pi,k_2) \qquad \forall k_2\in\T^1.$$
    \end{enumerate}
\end{definition}
    
Moreover, we will relate the lack of periodicity of the vector $v_1$ as above to the Chern number: $h =\Ch(P) \in \Z$.

\subsection{Chern number through the matching matrices}

Let $P:\T^2 \to \Proj_m(\Hi)$ be a continuous, not necessarily symmetric, projection-valued map, and let $U:[-\pi,\pi]\times \T^1 \to \U(\Hi)$ be a continuous unitary-valued map as in Proposition~\ref{pro:Kato-Nagy,d=2}, that is, periodic in the second argument and such that
    \[
        P(t,k_2) = U(t,k_2) P(0,0) U(t,k_2)^{-1} \quad \forall (t,k_2) \in [-\pi,\pi] \times \T^1.
    \]
Then $U(-\pi,k_2)^{-1} U(\pi,k_2)$ commutes with $P(0,0)$, so after choosing an orthonormal basis $\{v_j\}_{1 \le j \le m}$ spanning $\Imm P(0,0)$ we can define
    \[
        \alpha(k_2)_{a,b} := \langle v_a, U(-\pi,k_2)^{-1} U(\pi,k_2) v_b \rangle\,, \quad a,b \in \{1,\ldots,m\}.
    \]
The matrices $\alpha(k_2) = [\alpha(k_2)_{a,b}]_{1 \le a,b \le m} \in M_{m,m}(\C)$, called \emph{matching matrices}, are then unitary, and depend periodically and continuously on $k_2 \in \T^1$.

\begin{definition}[Chern number]\label{def:Chern_number}
If $m \in \N$, the \emph{Chern number} of the rank-$m$ projection-valued map $P$ is the integer $\Ch(P) \in \Z$ obtained as the \emph{winding number} of $\det \alpha \colon \T^1 \to \rU(1)$ (compare Definition~\ref{def:winding_number}):
    \[
        \Ch(P) := [\det(\alpha(\cdot))].
    \]
    If $m = \dim \Imm P(k) = \infty$, we set $\Ch(P) := 0$.
\end{definition}

In the remainder of this Section, we derive from the previous definition some fundamental properties of the Chern number. 

\begin{proposition}\label{pro:Ch_invariant}
    The above definition is independent of the choice of the unitary operator~$U$ and of the basis $\{v_j\}$ for $\Imm P(0,0)$. Moreover, if two projection-valued maps are Murray--von Neumann equivalent, unitarily equivalent or homotopic, their Chern numbers coincide.
\end{proposition}

\begin{proof}
\noindent \emph{Well-posedness}. \quad
First, we show that the Chern number does not depend on the choice of the orthonormal basis $\{v_j\}$. If $\{v_j'\}$ is another basis, it induces a different $\alpha'(k_2)$, which is unitarily equivalent to $\alpha(k_2)$ via $A_{a,b} := \langle v_a, v_b' \rangle$; consequently
\[
\alpha(k_2) = A^{-1} \alpha'(k_2) A \implies \det(\alpha(k_2)) = \det(\alpha'(k_2)).
\]
        Next, consider two unitary-valued maps $U, V:[-\pi,\pi]\times \T^1 \to \U(\Hi)$ satisfying
        \[
            P(t,k_2) = U(t,k_2) P(0,0) U(t,k_2)^{-1} = V(t,k_2) P(0,0) V(t,k_2)^{-1}.
        \]
        Then $V(t,k_2)^{-1} U(t,k_2)$ commutes with $P(0,0)$, and defining
        \[
            \beta_{a,b}(t,k_2) = \langle v_a, V(t,k_2)^{-1} U(t,k_2) v_b \rangle,
        \]
        continuity in $t$ and periodicity in $k_2$ imply that $[\det \beta(t,\cdot)]$ is constant in $t$. By Proposition \ref{cor:top_group} and $\det(AB) = \det(A)\det(B)$, we obtain
        \[
            [\det(\alpha(\cdot))] = [\det(V(-\pi,\cdot)^{-1} V(\pi,\cdot)|_{\Imm(P(0,0))})],
        \]
        proving independence of the Chern number on $U$.

\smallskip 

\noindent \emph{Murray--von Neumann and unitary invariance}. \quad Let $V(k)$ implement a Murray--von Neumann equivalence or a unitary equivalence between two projection-valued maps $P_0(k)$ and $P_1(k)$. In both cases we have
$$
P_1(k_1,k_2)= V(k_1,k_2)P(k_1,k_2)V(k_1,k_2)^*
$$

From proposition \ref{pro:Kato-Nagy,d=2}, there exists a unitary--valued map $U$, periodic in the second variable, such that
        \[
            P_0(t,k_2) = U(t,k_2) P_0(0,0) U(t,k_2)^{-1},
        \]
        Then:
        \[
            P_1(t,k_2) = V(t,k_2) U(t,k_2) V(0,0)^{*} P_1(0,0
            ) V(0,0) U(t,k_2)^{-1} V(t,k_2)^{*}.
        \]
        Choosing a basis $\{v_j^1\}$ of $\Imm(P_1(0,0))$ and defining
        \[
            \alpha_1(k_2)_{a,b} = \langle v_a^1, V(0,0) U(-\pi,k_2)^{-1} V(-\pi,k_2)^{*} V(\pi,k_2) U(\pi,k_2) V(0,0)^{*} v_b^1 \rangle,
        \]
        periodicity of $V$ in $k_2$ and the relation $v_j^0 := V(0,0)^{*} v_j^1$ lead to
        \[
            \alpha_1(k_2)_{a,b} = \langle v_a^0, U(-\pi,k_2)^{-1} U(\pi,k_2) v_b^0 \rangle.
        \]
        Hence,
        \[
            \Ch(P_1) = [\det(\alpha_1)] = [\det(\alpha_0)] = \Ch(P_0).
        \]

\smallskip

\noindent \emph{Homotopy invariance}. \quad Consider a homotopy $P_t(k)$ of projection-valued maps. By compactness of $[0,1]\times \T^2$ and continuity, we can partition $[0,1] = \bigcup_{j=0}^{J-1} [t_j, t_{j+1}]$ in a way such that
        \[
            \|P_{t_j}(k) - P_t(k)\| \le 1/2, \quad \forall t \in [t_j, t_{j+1}], k \in \T^2.
        \]
        Then, for each $j$, the Kato-Nagy unitary $U_j(k) := U_{P_{t_j}(k) \to P_{t_{j+1}}(k)}$ implements a unitary equivalence between $P_{t_j}$ and $P_{t_{j+1}}$. By the previous item, Chern numbers are preserved along each step, yielding
        \[
            \Ch(P_0) = \Ch(P_{t_1}) = \dots = \Ch(P_1). \qedhere
        \]
\end{proof}

\begin{proposition}\label{lem:additivity_Chern_Nuber}
    The Chern number is additive: if $P:\T^2 \to \Proj_r(\Hi)$ and $Q:\T^2 \to \Proj_m(\Hi)$ satisfy $PQ \equiv 0$, then $P+Q:\T^2 \to \Proj_{r+m}(\Hi)$ is still a continuous periodic projection-valued map satisfying
    \[
        \Ch(P+Q) = \Ch(P) + \Ch(Q).
    \]
\end{proposition}

\begin{proof} 
    Applying Proposition \ref{pro:Kato-Nagy,d=2} to $P$, $Q$, and $P+Q$, we obtain unitary equivalences $$U, V, Y:[0,2\pi]\times \T^1 \to \U(\Hi)$$
    such that
    \[
        P(t,k_2) = U(t,k_2) P(0,0) U(t,k_2)^{-1}, \]
        \[Q(t,k_2) = V(t,k_2) Q(0,0) V(t,k_2)^{-1},\]
        \[(P+Q)(t,k_2) = Y(t,k_2) (P+Q)(0,0) Y(t,k_2)^{-1}.
    \]
    We set \[ X(u) :=\begin{cases} U(u) \quad \mbox{if} \quad Pu=u \\
    V(u) \quad \mbox{if} \quad Qu = u \\
    Y(u) \quad \mbox{if} \quad Pu=Qu =0
    \end{cases} \]
     and extend the definition to $X:[0,2\pi]\times \T^1 \to \U(\Hi)$ by linearity, to obtain a unitary-valued map with \[(P+Q)(t,k_2) = X(t,k_2) (P+Q)(0,0) X(t,k_2)^{-1}.\]
    Now we compute the Chern number of $P+Q$ using a basis $\{ v_j\}_{j\in\{1,\cdots, r+m\}}$ where the first $r$ vectors span $P(0,0)$ and the other $m$ span $Q(0,0)$. In this way we have \[\alpha(k_2) = \begin{pmatrix}
        \beta(k_2) & 0  \\ 0 & \gamma(k_2)
    \end{pmatrix}\]
    where $\beta :\T^2 \to \rU(r)$ and $\gamma: \T^2 \to \rU(m)$ are such that
    \[
    \beta(k_2)_{a,b} = \inn{v_a}{U(-\pi,k_2)^{-1} U(\pi,k_2) v_b} \quad \text{and} \quad \gamma(k_2)_{a,b} = \inn{v_{r+a}}{V(-\pi,k_2)^{-1} V(\pi,k_2)v_{r+b}}\,.
    \]
    Applying Proposition \ref{cor:top_group} we obtain the desired equality: \[\Ch(P+Q) = [\det \alpha] = [\det\beta \cdot \det\gamma] = [\det\beta]+[\det\gamma] = \Ch(P) + \Ch(Q).\qedhere\]
\end{proof}

\subsection{Pseudo-periodic frames}

The next statement exhibits the construction of a pseudo-periodic frame for a continuous periodic projection-valued map, and expresses the Chern number as the obstruction to finding a fully periodic frame.

\begin{theorem}[Existence of a pseudo-periodic frame]\label{thm:pseudo-periodic_frame}
    Consider a projection-valued map $P :\T^2 \to \Proj_m(\Hi)$. Then it is always possible to construct a pseudo-periodic frame $\{v_j(k)\}_{1 \le j \le m}$ for $P$ as in Definition~\ref{def:pseudo-periodic frame}, with $h=\Ch(P) \in \Z$.
\end{theorem}

\begin{proof}    
    Using Proposition \ref{pro:Kato-Nagy,d=2} we can build a unitary-valued map periodic in $k_2$ such that \[P(t,k_2) = U(t,k_2) P(0,0) U(t,k_2)^{-1} \quad \mbox{for} \quad (t,k_2) \in [-\pi,\pi]\times \T^1.\]
    If $\{v_j\}_{1 \le j \le m}$ is an orthonormal basis of $\Imm P(0,0)$, then $\{u_j(t,k_2) = U(t,k_2)v_j\}_{1 \le j \le m}$ constitutes an orthonormal basis of $P(t,k_2)$ that depends continuously in $t$ and $k_2$. This is periodic in $k_2$ but in general will not be periodic in $t$. The aperiodicity is once again measured by the matching matrix $\alpha(k_2)\in \rU(m)$, with:\[u_j(\pi,k_2) = \sum_{a=1}^m [\alpha(k_2)]_{j,a} u_j(-\pi,k_2).\]
    Since $[\det \alpha]=\Ch(P)$ the matrix:\[\beta(k_2) = \begin{pmatrix}e^{i\Ch(P)k_2} & 0 \\ 0 & \Id_{n-1} \end{pmatrix} \alpha(k_2)^{-1}\] is such that $[\det\beta]=0$.
    By Theorem \ref{thm:homotopy_group_U(n)}, there exists a homotopy $\beta_s(k_2)$ with $\beta_1(k_2) = \beta(k_2)$ and $\beta_0(k_2)= \Id$.
    Therefore, the vectors\[v_j(t,k_2) = \sum_{a=1}^m\left[\beta_{(t+\pi)/2\pi} (k_2)\right]_{j,a} u_j(t,k_2) \quad \mbox{for} \quad j\in\{1,\cdots,n\}\]
    constitute an orthonormal basis of $\Imm P(t,k_2)$ and are such that:
    \begin{align*}
    v_j(\pi,k_2)& = \sum_{a=1}^m [\beta(k_2)]_{j,a}u_a(\pi,k_2) =\sum_{a,b=1}^m[\beta(k_2)]_{j,a} \alpha_{a,b}(k_2)u_b(-\pi,k_2) \\
    &= \sum_{a=1}^m [\beta(k_2) \alpha(k_2)]_{j,a}u_a(-\pi,k_2) = \begin{cases}
        e^{i\Ch(P)k_2}v_1(-\pi,k_2) &\mbox{for } j=1\,,\\
        v_j(-\pi,k_2) & \mbox{otherwise}. 
    \end{cases} \qedhere
    \end{align*}
\end{proof}

\section{From the symmetric matching matrices to the $\mathbb{Z}_2$-valued invariant}\label{section: The symmetric matching matrix}

Analogously to what happens in the absence of time-reversal symmetry, one can associate to a time-reversal symmetric projection-valued map on the torus a topological invariant, with values in $\Z_2$, obtained from its matching matrices. In particular, the fact that the projection-valued map is subject to time-reversal symmetry induces a corresponding symmetry on the associated matching matrices, provided these are computed from a pseudo-periodic and symmetric frame. This additional symmetry and its consequences are discussed in the first subsection, which essentially reviews some definitions and results contained in~\cite{Cornean_2017}. The second subsection defines the $\mathbb{Z}_2$ topological invariant, showing that it truly depends only on the family of projections and not on the choices made to define it from the matching matrices. Our definition can be related to the one proposed by Graf and Porta in~\cite{graf2013bulk}, as discussed in Remark~\ref{rmk:Graf Porta index}, and therefore, in turn, to other formulations of the invariant, including the original one by Fu, Kane and Mele \cite{Cornean_2017}.

\subsection{Review: time-reversal symmetric matching matrices} \label{sec:revTRS}

Consider a time-reversal symmetric projection-valued map $P:\T^2 \to \Proj_{2n}(\Hi)$ and a unitary-valued map $U:\T^2 \to \U(\Hi)$ ---obtained, for example, by using Proposition \ref{pro:Kato-Nagy,d=2}--- such that 
    \[P(t,k_2) U(t,k_2) = U(t,k_2) P(0,0) \quad \mbox{and} \quad TU(t,k_2) = U(-t,-k_2) T \quad \forall t\in[-\pi,\pi], k_2 \in \T^1.\] 
    Then we can use Lemma \ref{lem:anti-unitary_quaternionic_structure} to fix a quaternionic basis of $\Imm P(0,0)$, i.e., a basis compatible with the quaternionic structure induced by the time-reversal operator $T$: fix an orthonormal set $\{u_j\}_{j\in\{1,\cdots n\}}$, and  define $u_{n+j}=Tu_j$ for $j\in\{1,\cdots,n\}$ to have a complete basis of $\Imm P(0,0)$. Then we can get a continuous collection of orthonormal bases of $P(t,k_2)$ by setting, for all $t\in[-\pi,\pi]$, $k\in \T^1$, and $j\in\{1,\cdots, 2n\}$,
    \[u_j(t,k_2) = U(t,k_2) u_j  \quad \mbox{with}\quad Tu_j(t,k_2)=\sum_{a=1}^{2n}J_{j,a}u_a(-t,-k_2)\] 
    where $J$ is as in \eqref{eqn:J}. The lack of periodicity in $k_1 $ is expressed by a continuous family of matching matrices $\alpha(k_2) \in \rU(2n)$ such that \[u_j(\pi,k_2) = \sum_{a=1}^{2n} [\alpha(k_2)]_{j,a} u_a(-\pi,k_2).\]
    This family satisfies the symmetry constraint $J=\overline{\alpha(k_2)} J \alpha(-k_2)$: \[\begin{split}
        \sum_{a=1}^{2n} [J]_{j,a} u_a(-\pi,k_2) &=Tu_j(\pi,-k_2) = T\sum_{a=1}^{2n}[\alpha(k_2)]_{j,a} u(-\pi,-k_2) \\ &=\sum_{a,b=1}^{2n} \overline{[\alpha (k_2)]_{j,a}}[J]_{a,b} u_b(\pi,k_2)  \\
        &=\sum_{a,c=1}^{2n} \left[\overline{\alpha(k_2)}(J)\right]_{j,a} [\alpha (-k_2)]_{a,c} u_c(-\pi,k_2)  \\
        &=\sum_{a=1}^{2n}\left[\overline{\alpha(k_2)}(J) \alpha(-k_2)\right]_{j,a} u_a(-\pi,k_2)
    \end{split}\]
    
\begin{remark}\label{rmk:TRSPVM_have_trivial_Chern_number}
From the above symmetry constraint, we immediately deduce that, as stated in~\cite{panati2007triviality, monaco2015symmetry}, any $P:\T^2 \to \Proj_{2n}(\Hi)$ is a time-reversal symmetric projection-valued map has $\Ch(P)=0$. Indeed, we can compute the Chern number through a symmetric family of matching matrices with $J=\overline{\alpha(k_2)}J \alpha(-k_2)$.
This implies that $\det(\alpha(k_2))=\det(\alpha(-k_2))$, but a map with this property has a winding number equal to zero due to Lemma \ref{lem:triviality_of_reflective_map}.
\end{remark}
    
One natural question arises: how will the matching matrix change after the choice of a different symmetric and a frame $\{v_j(t,k_2)\}_{j\in\{1,\cdots , 2n\}}$ which is periodic in the second entry but not in the first? Clearly, there is a unitary-valued map $\beta:[-\pi,\pi]\times \T^1 \to \rU(2n)$ such that $v_j(t,k_2) = \sum_{a=1}^{2n} [\beta(t,k_2) ]_{j,a}u_a(t,k_2)$.
    The first condition on $\beta$ is that the second frame is symmetric if and only if $J\beta(t,k_2) = \overline{\beta (-t,-k_2)} J$: \[\begin{split}Tv_j(t,k_2) & =  \sum_{a=1}^{2n}[J]_{j,a} v_a(-t,-k_2) \Leftrightarrow \\
        \Leftrightarrow \sum_{a=1}^{2n}\overline{[\beta(t,k_2)]_{j,a}} T u_a(t,k_2)&=\sum_{a,b=1}^{2n}[J]_{j,a} [\beta(-t,-k_2)]_{a,b} u_b(-t,-k_2) \Leftrightarrow \\
        \Leftrightarrow \sum_{b=1}^{2n} \left[\overline{\beta(t,k_2)}(J)\right]_{j,b} u_b(-t,-k_2) & = \sum_{b=1}^{2n}\left[J\beta(-t,-k_2)  \right]_{j,b} u_b(-t,-k_2) \end{split}\]
        
    Moreover, if the new family of matching matrices is $\alpha'(k_2)$ with
    \[v_j(\pi,k_2) = \sum_{a=1}^{2n} [\alpha'(k_2)]_{j,a} v_a(-\pi,k_2),\]
    then $\alpha$ and $\alpha'$ are intertwined by the equation 
    \begin{equation}\label{eq:change_of_frame}
        \alpha'(k_2)=\beta(\pi,k_2)\alpha(k_2)\beta(-\pi,k_2)^{-1} \quad \forall k_2 \in \T^1.
    \end{equation}
    This is true because
    \[\begin{split}
            \sum_{a=1}^{2n}[\alpha'(k_2)]_{j,a}v_a(-\pi,k_2) & =v_j(\pi,k_2) \Leftrightarrow \\
            \sum_{a=1}^{2n}[\alpha'(k_2)\beta(-\pi,k_2)]_{j,a} u_a(-\pi,k_2) &= \sum_{a=1}^{2n}[\beta(\pi,k_2) ]_{j,a} u_a(\pi,k_2) \Leftrightarrow\\
            \sum_{a=1}^{2n}[\alpha'(k_2)\beta(-\pi,k_2)]_{j,a} u_a(-\pi,k_2) &= \sum_{a=1}^{2n}[\beta(\pi,k_2) \alpha(k_2) ]_{j,a} u_a(-\pi,k_2).\end{split} \]

We can interpret the previous conditions as follows: the collection of pseudo-periodic symmetric frames for a time-reversal symmetric projection-valued map is in bijection with the set $A / \sim$, where
\[
A :=\left\{\alpha:\T^1 \to \rU(2n) : J=\overline{\alpha(k_2)} J \alpha(-k_2)  \right\} 
\]
and where $\alpha\sim \alpha'$ in $A$ if and only if there is a map $\beta:[-\pi,\pi]\times \T^1 \to \rU(2n) $ such that 
\begin{equation}\label{eq:set_of_symmetric_matching_matrices}
J\beta(t,k_2) = \overline{\beta (-t,-k_2)}J \quad \mbox{and}\quad\alpha'(k_2)=\beta(\pi,k_2)\alpha(k_2)\beta(-\pi,k_2)^{-1} \quad \forall k_2 \in \T^1, t\in [-\pi,\pi].
\end{equation}

\begin{remark}\label{rmk:beta_has_even_winding_number}
The first of the above two conditions implies that the determinant of $\beta(t,\cdot)$ has an even winding number for all $t \in [-\pi,\pi]$.
    In fact, the symmetry condition $J\beta(0,k_2) = \overline{\beta (0,-k_2)} J$, forces $\beta(0,0), \beta(0,\pi)\in \rSp(2n)$, which in turn implies that $\det\beta(0,0) = \det\beta(0,\pi)=1$. Moreover, $\det \beta(0,k_2) = \overline{\det \beta(0,-k_2)}$, so we can use Lemma \ref{lem:even_winding_number} to conclude that the winding number in the second direction $[\det\beta(0,\cdot)]$ must be even. Continuity in $t$ ensures that $[\det \beta(t,\cdot)]$ is even (and constant) for every $t\in [-\pi,\pi]$.
\end{remark}

\subsection{The $\mathbb{Z}_2$-valued invariant}

The set of symmetric matching matrices has a topological invariant dating back to the milestone paper~\cite{kane2005z}, where Kane and Mele demonstrated that it is possible to associate a $\Z_2$-valued topological invariant to a specific discrete honeycomb model. 
The methods introduced in that work were subsequently extended to a broader class of models in several later papers, including~\cite{fu2007topological, freed2013twisted, graf2013bulk, de2015classification, fiorenza2016z, Prodan_2009}, each providing a distinct formulation of the invariant and a different interpretation of the physical information it encodes. 
In this subsection, elaborating on \cite{Cornean_2017}, we will present yet another formulation of the invariant, using the matching matrices formalism.

Consider a family of symmetric matching matrices $\alpha(k_2)$ such that $J=\overline{\alpha(k_2)}J \alpha(-k_2)$ for all $k_2 \in \T^1$. Using this symmetry constraint we can find two unitary matrices $\gamma_0, \gamma_\pi$ such that 
\[ \alpha(0) = J \gamma_0^t J^{-1} \gamma_0 \]
and similarly for $\gamma_\pi$. For example, since $J \alpha(0)^t J^{-1} = \alpha(0)$, we can find a self-adjoint matrix $L$ such that $\alpha (0) = e^{iL}$ and $JL^tJ^{-1} = L$: then $\gamma_0 = e^{iL/2}$ will work. It is obvious that $\det \alpha(0) = (\det \gamma_0)^2$ and $\det \alpha(\pi)=(\det \gamma_\pi)^2$. If $\det \gamma_0=e^{i\lambda_0}$ and $\det\gamma_\pi=e^{i\lambda_\pi}$ for two real numbers $\lambda_0, \lambda_\pi\in \R$, we can choose a continuous function $\mu:[0,\pi]\to \R$ such that $\det \alpha(k_2) = e^{i\mu(k_2)}$ and $\mu(0) = 2\lambda_0$. Then it holds that $e^{i\mu(\pi)} = e^{i2\lambda_\pi}$, so $\mu(\pi)-2\lambda_\pi\in 2\pi \Z$.

\begin{definition}[$\delta(P)$]\label{def:Parity_invariant}
  Under these hypotheses, we will define:\[\delta (P)= \delta(\alpha):=e^{i\frac{\left(2\lambda_\pi-\mu(\pi)\right)}{2}} \in \{\pm 1\}= \Z_2.\]
\end{definition}

\begin{proposition}\label{pro:Parity_invariant_well_posed}
    The quantity defined previously does not depend on the choices of $\gamma_0, \gamma_\pi \in \rU(2n)$, of $\lambda_0, \lambda_\pi \in \R$, and is also well-posed with respect to the equivalence relation in~\eqref{eq:set_of_symmetric_matching_matrices}, meaning that $\delta(\alpha) = \delta(\alpha')$ if $\alpha \sim \alpha'$ in $A$. Therefore it depends only the symmetric projection-valued map $P$, allowing for the abuse of notation $\delta(P)=\delta(\alpha)$.
\end{proposition}

\begin{proof}
    If we took $\lambda'(0)$, $\lambda'(\pi)$ such that $e^{i\lambda_0}=e^{i\lambda'(0)}$ and $e^{i\lambda_\pi}=e^{i\lambda'(\pi)}$, we would have that $\lambda (0) = \lambda'(0) + 2\pi l$, $\lambda_\pi = \lambda'(\pi)+2\pi m$ for some $l,m \in \Z$. This means that we need to impose $\mu'(k) = \mu(k)+4\pi$ and obtain: \[\delta' = e^{i\frac{2\lambda'(\pi)-\mu'(\pi)}{2}}  = e^{i\frac{2\lambda_\pi + 4\pi m-\mu(\pi)-4\pi l}{2}}=\delta\cdot e^{i2\pi(m-l)}=\delta\]
    Instead if we took another $\gamma'_0$ such that $\alpha(0) = J \gamma_0^t J^{-1}\gamma_0 = J (\gamma'_0)^t J^{-1}\gamma'_0$, then it is true that \[(\gamma'_0)^t J\gamma'_0 = \gamma_0^t J \gamma_0 \Leftrightarrow J = (\gamma_0(\gamma_0')^{-1})^t J (\gamma_0 (\gamma_0') ^{-1}).\] This means that $\gamma_0(\gamma'_0)^{-1} \in \rSp (2n)$, so $\det(\gamma_0 (\gamma'_0)^{-1})=1$ and therefore $\det(\gamma_0) = \det (\gamma'_0)$. Since the same holds for $\gamma_\pi$, we have that $\det(\gamma_0), \det(\gamma_\pi)$ depends only on $\alpha(0), \alpha(\pi)$.
    
    Instead, suppose that we want to study the invariant of the family \[\alpha'(k_2) = \beta(\pi,k_2)\alpha(k_2)\beta(-\pi,k_2)^{-1} \quad \mbox{for } \beta \mbox{ such that } \quad J\beta(t,k_2)=\overline{\beta (-t,-k_2)} J.\]
    Then we can use the symmetry constraint on $\beta$ to rewrite
    \[\alpha'(k_2) = \beta(\pi,k_2)\alpha(k_2)J^{-1}\beta(\pi,-k_2)^{t} J\]
    which shows how we can set 
    \[ \gamma_0':=\gamma_0 J^{-1}\beta(\pi,0)^tJ \quad \text{and} \quad \gamma_\pi':=\gamma_\pi J^{-1}\beta(\pi,\pi)^tJ \]
    and obtain $\alpha'(0)=J(\gamma_0')^t J^{-1}\gamma_0'$ and $\alpha'(\pi)=J(\gamma_\pi')^t J^{-1}\gamma_\pi'$. 
    Now, given a continuous function $b:[-\pi,\pi]^2\to \R$ such that $\det\beta(t_1,t_2)=e^{ib(t_1,t_2)}$, we have that $\lambda_0'=\lambda_0+b(0,0)$, $\lambda_\pi'=\lambda_\pi+b(\pi,0)$ and $\mu'(k_2)=\mu(k_2)+b(\pi,k_2)+b(\pi,-k_2)$.
    Since $2\lambda_0'=2\lambda_0+2b(\pi,0)=\mu'(0)$, we can compute the invariant by studying \[2\lambda_\pi'-\mu'(\pi)=2\lambda_\pi+2b(\pi,\pi) -\mu(\pi)-b(\pi,\pi)-b(\pi,-\pi)=2\lambda_\pi-\mu(\pi)+b(\pi,\pi)-b(\pi,-\pi).\]
    According to Definition \ref{def:winding_number}, $[b(\pi,\pi)-b(\pi,-\pi)]/2\pi$ is the winding number of $[\det\beta(\pi,\cdot)]$ in the second direction. In view of Remark \ref{rmk:beta_has_even_winding_number}, this is even, so $b(\pi,\pi)-b(\pi,-\pi) \in 4\pi \Z$. We conclude that that
    \[e^{i\frac{2\lambda_\pi'-\mu'(\pi)}{2}} = e^{i\frac{2\lambda_\pi-\mu(\pi)}{2}}e^{i\frac{b(\pi,\pi)-b(\pi,-\pi)}{2}} = e^{i\frac{2\lambda_\pi-\mu(\pi)}{2}}.\]
    Therefore, the invariant is also well-posed with respect to the equivalence relation \eqref{eq:set_of_symmetric_matching_matrices}, as claimed.
\end{proof}

\begin{remark}\label{rmk:Graf Porta index}
    The $\mathbb{Z}_2$ invariant introduced above can be viewed as a reformulation of the Graf--Porta invariant, first introduced in~\cite{graf2013bulk}. This reformulation is analogous to the one contained in Proposition 5.3 in~\cite{Cornean_2017}, but extended to the case in which the projection-valued map $P$ has regularity $C^0$ rather than $C^1$. The above-mentioned Proposition shows that the Graf--Porta invariant $\mathcal{I}(\alpha)$ of a matching matrix $\alpha$ can be computed as the sum of twice the difference between the Pfaffians of $J\alpha(\pi)$ and $J\alpha(0)$ (which are anti-symmetric matrices), and the ``partial winding number'' of the determinant of the matching matrix $\alpha$ restricted to $[0,\pi]$. This winding number is ``partial'' because on $[0,\pi]$ the map $\alpha$ does not define a closed loop; however, in the $C^1$ setting it can still be computed as the integral, restricted to $[0,\pi]$, of the derivative of $\log \det \alpha(k_2)$.

    In the $C^0$ setting, this derivative is not available, but the partial winding number can nonetheless be defined as the difference between the initial and final values of any continuous lift to the universal cover $\mathbb{R}$ of $\rU(1)$ of the map $\det \alpha$ restricted to $[0,\pi]$. Such a lift is simply a continuous logarithm of $\det \alpha(k_2)$, since the covering map from $\mathbb{R}$ to $\rU(1)$ is the exponential. In this context, the map $\mu(k_2)$ defined earlier provides exactly such a logarithm, while the values $\lambda_0$ and $\lambda_\pi$ are the Pfaffians of $J\alpha(0)$ and $J\alpha(\pi)$, respectively. Since $\mu(0)=2\lambda_0$, we obtain
    \[
        2\lambda_\pi-\mu(\pi)
        = 2(\lambda_\pi-\lambda_0)\;-\;(\mu(\pi)-\mu(0))
        = 2\Pf(J\alpha(\cdot))\big|_0^\pi\;+\;\log\det\alpha(\cdot)\big|_0^\pi.
    \]
    This, together with the discussion in~\cite{Cornean_2017}, shows that the invariant defined in this work reproduces the Graf--Porta Index.
\end{remark}

Using the invariant, the following theorem states that $A/\!\sim$ has only two elements, given by the equivalence classes identified by $\delta=1$ and $\delta=-1$. This result already appears in~\cite{Cornean_2017}, but we include the proof ---adapted to the present formalism--- for completeness.

\begin{theorem}\label{thm:connecting_mathcing_matrices_after_frames_changing}
    Given two families of symmetric matching matrices $\alpha, \alpha' \in A$, the $\alpha \sim \alpha'$ as in \eqref{eq:set_of_symmetric_matching_matrices} if and only if $\delta (\alpha)=\delta(\alpha')$.
\end{theorem}

\begin{proof}
The previous result already showed the constancy of $\delta$ on $\sim$-equivalence classes, so we need to show that, conversely, if $\delta(\alpha) = \delta(\alpha') \in \Z_2$ then it is possible to construct a function $\beta:[-\pi,\pi]\times S^1 \to \rU(2n)$ such that 
\begin{equation}\label{eq:conditions_in_splitting_theroem}
\beta(t,k_2)=J^{-1}\overline{\beta(-t,-k_2)}J \mbox{ and } \alpha'(k_2)= \beta (\pi,k_2) \alpha(k_2) J^{-1}\beta (\pi,-k_2)^tJ. 
\end{equation}
At $(t,k_2) = (\pi,0)$ the equation on the right reduces to $ \alpha'(0)=\beta(\pi,0)\alpha(0)J^{-1}\beta(\pi,0)^tJ$, which is satisfied for $\beta(\pi,0):=\gamma_0'\gamma_0^{-1}$; the same holds at $(t,k_2) = (\pi,-\pi)$ for $\beta(\pi,\pi):=\gamma_\pi'\gamma_\pi^{-1}$. 
Since all $\gamma$'s can be chosen to be symplectic matrices, we have $\beta(\pi,0),\beta(\pi,\pi) \in \rSp(2n)$.

The next step is to extend the definition of $\beta$ to the segment $\overline{(\pi,-\pi)(\pi,0)}$. This can be achieved by choosing any continuous path $p \colon [0,1] \to \rU(2n)$ with $p(0)=\beta(\pi,0)$ and $p(1)=\beta(\pi,\pi)$, which exists since $\rU(2n)$ is path-connected, allowing us to define $\beta(\pi,k_2)=p(-k_2/\pi)$ for $k_2\in[-\pi,0]$. To further extend the definition for $k_2\in[0,\pi]$, we impose the second condition in~\eqref{eq:conditions_in_splitting_theroem} and define
\[\beta(\pi,s)=\alpha'(s)J^{-1}\overline{\beta(\pi,-s)}J\alpha(s)^{-1} = \alpha'(s) J^{-1}\overline{p(-s/\pi)}J \alpha(s)^{-1} \quad \forall s \in [0,\pi].\]

Before moving on, we need to check that the winding number of $\det\beta(\pi,k_2)$ is even, in order not to violate the condition in Remark \ref{rmk:beta_has_even_winding_number}.
Using the same notation we used when we computed $\delta(\alpha')$, we have that $\det\beta(\pi,0)=1=e^{i(\lambda_0'-\lambda_0)}$ and $\det\beta(\pi,\pi)=1=e^{i(\lambda_\pi'-\lambda_\pi)}$.
Moreover, it is possible to chose a function $l:[0,1] \to \R$ such that $\det(p(s))=e^{il(s)}$ with $l(0)=\lambda_0'-\lambda_0, l(1)-\lambda_\pi'+\lambda_\pi \in 2\pi \Z$ and use it to express
\[\det\beta(\pi,k_2)=\begin{cases}
    e^{il(- k_2/\pi)} & \mbox{if } k_2 \in [-\pi,0], \\
    e^{i[\mu'(k_2)-l(k_2/\pi)-\mu(k_2)]} & \mbox{if } k_2 \in [0,\pi],
\end{cases}\]
and finally $\det\beta(\pi,k_2)=e^{ib(k_2)}$ where
\[b(k_2)=\begin{cases}
    l(-k_2/\pi) & \mbox{if } k_2 \in [-\pi,0], \\
    \mu'(k_2)-l( k_2/\pi)-\mu(k_2) & \mbox{if } k_2 \in [0,\pi].
\end{cases}\] 
The winding number of $\det(\beta(\pi,k_2))$  is now
\[[\det\beta(\pi,\cdot)]=\frac{\mu'(\pi)-\mu(\pi)-l(1)-l(1)}{2\pi}.\]
However $2l(1) \equiv -2 \lambda_\pi'+2\lambda_\pi  \bmod 4\pi$, so
\[[\det(\beta(\pi,\cdot))]\equiv\frac{\mu'(\pi)-2\lambda_\pi'-\mu(\pi)+2\lambda_\pi}{2\pi} \bmod 2.\]
But this quantity is even because
\[e^{i\frac{\mu'(\pi)-2\lambda_\pi'-\mu(\pi)+2\lambda_\pi}{2}} = \delta(\alpha')\delta(\alpha)=1.\]

All these arguments ensure that the second condition of Equation \eqref{eq:conditions_in_splitting_theroem} holds, so we move to the discussion of the first symmetry condition.
At $t=0$, it reads $J\beta(0,k_2)=\overline{\beta(0,-k_2)}J$.
If $[\det(\beta(\pi,\cdot))]=2r$, then we can define \[\beta(0,k_2)  := \diag (e^{irk_2}, \underbrace{1,\cdots,1}_{n-1} , e^{irk_2}, \underbrace{1,\cdots, 1}_{n-1})\] and it is trivial to check that $\beta(0,k_2) $ satisfies the symmetry constraint and has $[\det\beta(0,\cdot)] = 2r = [\det\beta(\pi,\cdot)] $.
 By \ref{thm:homotopy_group_U(n)} we can find a homotopy $\beta_t(k_2)$ in $\rU(2n)$ with $\beta_0(k_2) = \beta(0,k_2)$, $\beta_1(k_2)  =\beta(\pi,k_2)$ and then define \[\beta(t,k_2) := \begin{cases}
            \beta_{t/\pi}(k_2) &\mbox{for } t\in[0,\pi]\\
            J^{-1} \overline{\beta_{-t/\pi}(-k_2)} J & \mbox{for } t\in [-\pi,0]
        \end{cases}\]
This map satisfies all the desired conditions and completes the proof of the Theorem.
\end{proof}

\section{Results}

In this Section, we present some features of a time-reversal symmetric projection-valued map $P$ on $\T^2$ which can be deduced from our definition \ref{def:Parity_invariant} of the invariant $\delta(P) \in \Z_2$.

\subsection{Symmetric splitting and pseudo-periodic symmetric frame}

Through the next theorem, which constitutes the main result of this text, we relate the possible choices of pseudo-periodic symmetric frame for $P$ to choices of a symmetric splitting for it, in the sense of Definition~\ref{def:symmetric_splitting}.

\begin{theorem}\label{thm:equivalence_theorem} 
    Consider a time-reversal symmetric projection-valued map $P:\T^2 \to \Proj_{2n}(\Hi)$ and a pseudo-periodic symmetric frame $\{u_j(t,k_2)\}_{j\in \{1,\cdots, 2n\}}$ for it, with associated family of matching matrices $\alpha \colon \T^1 \to \rU(2n)$. For $h \in \Z$, the following statements are equivalent:
    \begin{enumerate}
        \item There exists a time-reversal symmetric splitting of $P(k)$ in the sense of Definition \ref{def:symmetric_splitting} with $\Ch(P^-)=h$.
        \item There exists a pseudo-periodic symmetric frame $\{v_j(t,k_2)\}_{j\in \{1,\cdots, 2n\}}$ in the sense of Definition \ref{def:pseudo-periodic symmetric frame} with pseudo-periodicity 
        \[ v_1(\pi,k_2)=e^{ihk_2}v_1(-\pi,k_2), \qquad v_{n+1}(\pi,k_2)=e^{-ihk_2}v_{n+1}(-\pi,k_2). \]
        \item The matching matrices $\alpha$ are $\sim$-equivalent in $A$ to
        \[\alpha'(k_2) := \diag (e^{ihk_2},\underbrace{1,\cdots, 1}_{n-1}, e^{-ihk_2},\underbrace{1,\cdots, 1}_{n-1})\]
        (compare \eqref{eq:set_of_symmetric_matching_matrices}). 
        \item $\delta (P)=(-1)^h \in \Z_2$.
    \end{enumerate}
\end{theorem}

\begin{proof}
    The first statement implies the second because, given a splitting $P(k)=P^+(k)+P^-(k)$, we can apply Theorem \ref{thm:pseudo-periodic_frame} to construct a pseudo-periodic frame $\{v_j(t,k_2)\}_{j\in\{1,\cdots,n\}}$ of $P^-$, in the sense of Definition \ref{def:pseudo-periodic frame}, with the first vector satisfying $v_1(\pi,k_2)=e^{i \Ch(P^-)k_2} v_1(-\pi,k_2)$. Then we can then set $v_{n+j}(t,k_2):=Tv_j(-t,-k_2)$ and define a pseudo-periodic frame for $P^+$. The union of the two collections will constitute a pseudo-periodic symmetric frame of $P$ in the sense of Definition~\ref{def:pseudo-periodic symmetric frame}.

    Conversely, the second statement implies the first because, given such a pseudo-periodic frame, we can define \[P^-(t,k_2) := \sum_{1 \le j \le n} \left| v_j(t,k_2) \right\rangle \left \langle v_j(t,k_2) \right|, \quad P^+(t,k_2) := \sum_{n+1 \le j \le 2n} \left| v_j(t,k_2) \right\rangle \left \langle v_j(t,k_2) \right|.\]
    The symmetry condition on the frame imposes $TP^+(t,k_2)=P^-(-t,-k_2)T$ and it is obvious that $P^+(t,k_2)P^-(t,k_2)\equiv 0$. Moreover, the pseudo-periodic condition still implies 
    \[\left| v_j(-\pi,k_2) \right\rangle \left \langle v_j(-\pi,k_2) \right| = \left| v_j(\pi,k_2) \right\rangle \left \langle v_j(\pi,k_2) \right| \] 
    so we can interpret $P^\pm$ as functions of $k \in \T^2$. Moreover, the family of matching matrices associated to the frame $\{v_j(-\pi,k_2)\}_{1 \le j \le n}$ for $P^-$ is $\tilde{\alpha}(k_2):=\diag(e^{ihk_2},1,\ldots,1) \in \rU(n)$, leading to $\Ch(P^-)=h$ according to Definition \ref{def:Chern_number}.

   The equivalence between the second and the third statements follows the discussion in Section~\ref{sec:revTRS}.
   
    Finally, the equivalence between the third and the fourth statements follows naturally from Theorem \ref{thm:connecting_mathcing_matrices_after_frames_changing} and the computation of $\delta(\alpha')$.
    In fact, it is immediate to check that $\det \alpha'(k_2)\equiv 1$, so we can take $\mu'(k_2)\equiv 0$; moreover, from $\alpha'(0)=\Id$ we can chose $\gamma_0'=\Id$ and $\lambda_0'=0$.
From 
\[\alpha'(\pi)= \diag ((-1)^h , \underbrace{1,\cdots,1}_{n-1} , (-1)^h, \underbrace{1, \cdots, 1}_{n-1}),\]
it is possible to chose
\[\gamma'_\pi=  \diag (i^h , \underbrace{1,\cdots,1}_{n-1} , i^h, \underbrace{1, \cdots, 1}_{n-1}).\]
Since $\det(\gamma_\pi')=(-1)^h$, we can choose $\lambda_\pi' = h$.
All these choices lead to $\delta(\alpha')=(-1)^h$ and so we can conclude the proof of the Theorem.
\end{proof}

\subsection{Addivity and stability of the $\Z_2$ invariant}

The main consequence of Theorem \ref{thm:equivalence_theorem} is that the $\mathbb{Z}_2$ invariant of a projection-valued map $P$ can be expressed through the Chern number of a projection-valued map $P^{-}$ for any symmetric splitting $P = P^{+} + P^{-}$. This relation between the two invariants is not new in the computational physics literature, and has been used to construct local markers for the Kane–Mele invariant \cite{Gilardoni_2022, Ba__2024}. As already observed in~\cite{Ba__2024}, the stability of such local markers depends on the spatial localization of the symmetric splitting, that is, in the present formalism, on the regularity of the associated projection-valued map. The previous result not only provides theoretical support for this numerical procedure, but also yields a method for constructing a sufficiently regular symmetric splitting.

Besides, Theorem \ref{thm:equivalence_theorem} allows us to directly prove two fundamental properties of $\delta$: additivity and invariance with respect to the equivalence relations introduced in Definition \ref{def:Equivalence relations for time-reversal symmetric projection-valued maps}. Both properties are already well known in the literature for other formulations of the $\Z_2$ invariant, as they arise naturally in geometric or $K$-theoretic formulations (e.g.~\cite{de2012topological, kellendonk2019cycliccohomologygradedcralgebras}), or can be proved explicitly in analytic formulations analogous to the one presented in this article \cite{Cornean_2017}.

\begin{proposition}\label{pro:delta_unitarily_homotopic_invariant} 
    If two time-reversal symmetric projection-valued maps $P_0,P_1:\T^2 \to \Proj_{2n}(\Hi)$ are Murray--von Neumann equivalent, unitarily equivalent or homotopic (in the sense of definition \ref{def:TRSPVM}), then $\delta(P_0)=\delta(P_1)$.
\end{proposition}

\begin{proof}
    Suppose that there is a Murray--von Neumann equivalence $V$ or a unitary equivalence $U$. Through Theorem \ref{thm:equivalence_theorem} we can obtain symmetric splitting $P_0=P_0^-(k) + P_0^+(k)$ with $P_0^\pm : \T^2 \to \Proj_n(\Hi)$ and $TP_0^+(k) = P_0^-(-k)T$. Clearly 
    \[P_1(k) = V(k)P_0^-(k) V(k)^* + V(k)P_0^+(k)V(k)^{*}\]
    or
    \[P_1(k) = U(k)P_0^-(k) U(k)^{-1} + U(k)P_0^+(k)U(k)^{-1}\] 
    are valid splittings of $P_1(k)$ in the sense of Definition \ref{def:symmetric_splitting}. Since the Chern number is stable under Murray--von Neumann and unitary equivalences (Proposition \ref{pro:Ch_invariant}), we have that $\Ch(P_0^-)=\Ch(P_1^-)$.
    So, it follows that: \[\delta(P_1) = e^{i\pi \Ch(P_1^-) }= e^{i\pi \Ch(P_0^-)} = \delta(P_0).\]
    
    Instead, suppose that there is an homotopy $P_t(k)$ with $TP_t(k) = P_t(-k)T$ for $t\in [0,1]$ that connects $P_0$ with $P_1$. Then we can divide $[0,1]$ into a partition $[t_j,t_{j+1}]$, ${j\in\{0,\cdots,J-1\}}$, such that $t_0=0, t_J=1$ and such that $\|P_t(k)-P_{t_j}(k)\| <1$ for all $t\in [t_j,t_{j+1}]$, $k\in \T^2$. Then we can use the Kato-Nagy formula \eqref{eq:Kato-Nagy_unitary_equivalence} to define the unitary-valued maps $U_j(k) := U_{P_{t_j}(k) \to P_{t_{j+1}}(k)}$ for all $j\in\{0,\cdots, J-1 \}$. It is easy to see that $U_j(k)$ is a symmetric unitary equivalence between $P_{t_j}(k)$ and $P_{t_{j+1}}(k)$ for all $j\in\{0,\cdots, J-1\}$, so, using the previous point, we see that $\delta$ is preserved at each step and therefore $\delta(P_0)=\delta(P_1)$.
\end{proof}

\begin{lemma}\label{lem:delta_additive}
    Consider two time-reversal symmetric projection-valued maps $P:\T^2 \to \Proj_{2n}(\Hi)$, $Q:\T^2 \to \Proj_{2m}(\Hi)$ such that $P(k)Q(k)\equiv 0$: then $\delta(P)\delta(Q) = \delta(P+ Q)$.
\end{lemma}

\begin{proof}
    We can use Theorem \ref{thm:equivalence_theorem} to find two symmetric splittings $P(k) = P^+(k) + P^-(k)$, $Q(k) = Q^+(k)+ Q^-(k)$. 
    Then it is obvious that \[P(k)+ Q(k) = [P^-(k)+ Q^-(k)] + [P^+(k) + Q^+(k)] \]
    defines a symmetric splitting of $P+Q$. From Lemma \ref{lem:additivity_Chern_Nuber} we have $\Ch(P^- + Q^-) = \Ch(P^-)+\Ch(Q^-)$, so we can conclude using \ref{thm:equivalence_theorem} that
    \[\delta(P+ Q)= e^{i\pi \Ch(P^-+ Q^-)} = e^{i\pi\Ch(P^-)}\cdot e^{i\pi \Ch(Q^-)}= \delta(P) \cdot \delta(Q).\qedhere\]
\end{proof}

\subsection{From pseudo-periodic frames to periodic Parseval frames} \label{sec:parseval}

Theorem \ref{thm:equivalence_theorem} yields the following conclusion: every continuous time-reversal symmetric projection-valued map $P$ splits as a sum of two orthogonal continuous time-reversal symmetric maps
\[ P(k) = P_2(k) + P_{2(n-1)}(k)\,, \quad P_2(k) \, P_{2(n-1)}(k) \equiv 0\,, \]
where $P_2(k)$ is the orthogonal projection on the $2$-dimensional subspace spanned by the Kramer pair of vectors in a pseudo-periodic symmetric frame which are not periodic, and $P_{2(n-1)}(k) = P(k) - P_2(k)$ is the projection onto the $(2n-2)$-dimensional subspace spanned by the periodic vectors in the frame. Moreover, by the same result, we have that $\delta(P) = \delta(P_2) \in \Z_2$, and that $\delta(P_{2(n-1)}) = 0 \in \Z_2$ (as $P_{2(n-1)}$ admits a continuous, periodic and symmetric frame).

Consider now the projection-valued map
\[ Q(k) := P_2(k) \oplus P_2(k) \quad \text{on } \cH \oplus \cH\,. \]
It is clear that $Q(k)$ is a rank-4 projection-valued map which is continuous, periodic, and time-reversal symmetric with respect to the time-reversal symmetry operator $T \oplus T$ on $\cH \oplus \cH$. Moreover, by additivity of the $\Z_2$ invariant (Lemma~\ref{lem:delta_additive}), we have
\[ \delta(Q) = \delta(P_2) \, \delta(P_2) = 1 \bmod 2 \]
and therefore, by Theorem \ref{thm:equivalence_theorem}, it admits a periodic symmetric frame 
\[ \{u_1(k), u_2(k), u_3(k) = T u_1(-k), u_4(k) = T u_2(-k)\} . \]
Call $w_j(k) := \Pi( u_j (k))$, $1 \le j \le 4$, where $\Pi \colon \cH \oplus \cH \to \cH$ is the projection on the first summand in the direct sum. While not orthonormal, the vectors $\{w_1(k), \ldots, w_4(k)\}$ are still continuous, periodic, time-reversal symmetric, and span $\Imm P_2(k)$; moreover, the \emph{Parseval condition} holds, namely
\[ \psi = \sum_{j=1}^{4} \inn{w_j(k)}{\psi} \, w_j(k) \quad \text{for all } \psi \in \Imm P_2(k)\,. \]

Following the lines of~\cite{cornean2019parseval} (compare also \cite{auckley2018parseval}) we have thus proved that
\begin{theorem} 
Any continuous, periodic and time-reversal symmetric rank-$2n$ projection-valued map can be spanned by a \emph{Parseval frame} consisting of $2n+2$ continuous, periodic and symmetric vectors, namely of $n+1$ continuous and periodic Kramers pairs.
\end{theorem}

\subsection{Emerging spin--type simmetry}

Let $P$ be a time–reversal symmetric projection–valued map, and consider a continuous unitary–valued map $S:\T^2\to U(\Hi)$ that squares to the identity, commutes with $P$, and satisfies the following compatibility relation with the time-reversal symmetry $T$:
\[
T S(k) = -S(-k)T, \qquad \forall k \in \mathbb{T}^2.
\]
We will refer to such a map as a \emph{spin-type symmetry} of $P$. Given a spin-type symmetry $S$, the Spin-Chern number of $P$ is defined as the integer-valued quantity
\[
\mathrm{Ch}_S(P) := \frac{\Ch\!\left(P \frac{S+\Id}{2}\right) - \Ch\!\left(P \frac{S-\Id}{2}\right)}{2}.
\]
Indeed, since $S^2=\Id$, the operators $(S\pm\Id)/2$ define continuous projection-valued maps. Moreover, since $[S,P]=0$, the operators $P^\pm:=P(S\pm\Id)/2$ are themselves projection-valued, and hence their Chern numbers $\Ch(P^\pm)$ are well defined. Since $(S(k)\pm\Id)/2$ are the spectral projections onto the eigenspaces of $S(k)$, the decomposition $P=P^++P^-$ is orthogonal. By Lemma~\ref{lem:additivity_Chern_Nuber}, one then has
\[
0=\Ch(P)=\Ch(P^+)+\Ch(P^-),
\]
which implies that $\Ch(P^+)$ and $\Ch(P^-)$ coincide modulo $2$, so that their half-difference $\Ch_S(P)$ is an integer. Furthermore, the same relation yields
\begin{equation}\label{eq:Spin-Chern formula}
\Ch_S(P)=\Ch(P^\pm)=\Ch\bigg(P\frac{S\pm\Id}{2}\bigg)\mod 2.
\end{equation}

Note that we have not assumed that $S$ is independent of $k$, although this is typically the case in physical models where the Spin-Chern number is defined. In fact, the operator $S$ usually corresponds to a conserved spin component and acts as an internal symmetry of the system, that is, it acts only on the second factor of the tensor product $L^2(X)\otimes \mathbb{C}^n$ on which the Hamiltonian is defined. When such an operator is decomposed through the Bloch--Floquet--Zak transform, its fiber representation is always independent of the quasi-momentum variable~$k$, since the transform acts only on the first factor of the tensor product. 

Clearly, the independence of $S$ from $k$ is not required in order to define the associated Spin-Chern number. In what follows, we show that for any time-reversal symmetric projection-valued map $P$ it is always possible to construct a spin-type symmetry $S$, and that the $\mathbb{Z}_2$~invariant of $P$ can be computed through the associated Spin-Chern number as
\[
\delta(P)=(-1)^{\Ch_S(P)}.
\]
In general, such an operator $S$ will be $k$-dependent and not unique; however, the Spin-Chern number of $P$ depends on the choice of $S$ only modulo $2$.

It is worth noting that the idea of computing $\mathbb{Z}_2$-valued invariants via additional symmetries has appeared previously in the literature. In \cite{kellendonk2019cycliccohomologygradedcralgebras}, a highly general framework based on Van~Daele $K$-theory and cyclic cohomology is developed, where torsion-valued pairings give rise to $\mathbb{Z}_2$ topological invariants. In Section~7.3 of that work, this general theory is applied to tight-binding, possibly non-periodic systems subject to fermionic time-reversal symmetry. It is shown that any such system admits, up to stabilization, a representative of its $K$-theory class carrying an additional spin-type symmetry, and that the resulting torsion-valued pairing, which expresses the expected $\mathbb{Z}_2$ invariant, reduces to its Spin-Chern number associated with the additional symmetry. The content of this section may therefore be viewed as a direct, non-$K$-theoretic analogue of these results in the periodic, not necessarily tight-binding, setting. In particular, we emphasize that the construction of the spin-type symmetry presented below requires no stabilization of the projection-valued map $P$.

Before proving the main theorem, we introduce the following technical lemma, which essentially allows one to construct symmetric, continuous, and periodic frames for infinite-dimensional time-reversal symmetric projection-valued maps.

\begin{lemma}\label{lem:infinite_dim_Class_AII_frame_d=2}
    If $\Hi$ is a separable Hilbert space with $\dim(\Hi) = \infty$ and $P:\T^2 \to \Proj_{\infty}(\Hi)$ is a continuous time-reversal symmetric projection-valued map with $TP(k)=P(-k)T$, then there is always a continuous and periodic collection of orthonormal vectors $\{v_j(k)\}_{j\in\N}$ such that \[\{v_1(k), -Tv_1(-k), \cdots, v_j(k),- Tv_j(-k), \cdots \}\] is an orthonormal basis of $\ker(P(k))$.
\end{lemma}

\begin{proof}
    We can start by using Lemma \ref{lem:anti-unitary_quaternionic_structure} to select a discrete quaternionic basis $\{v_j\}_{j\in \N}$ of $\ker P(0,0)$. This basis, together with $\{Tv_j\}_{j\in \N}$, is a complex basis of $\ker P(0,0)$. Then we can use Proposition \ref{pro:Kato-Nagy,d=2} to construct a unitary-valued map $U(t,k_2)$ such that \[P(t,k_2) = U(t,k_2) P(0,0) U(t,k_2)^{-1} \quad \mbox{and} \quad T U(t,k_2) = U(-t,-k_2) T.\]
    We can construct an orthonormal basis of $\ker P(t,k_2)$ using the vectors \[\{U(t,k_2)v_1, U(t,k_2)Tv_1, \cdots, U(t,k_2) v_j ,U(t,k_2) Tv_j, \cdots\}.\] 
    This is periodic in the second argument, but not in the first. To solve this problem, we notice that this family defines a family of natural isomorphisms: \[ \begin{matrix}
        I(t,k_2): & \ker(P(t,k_2)) & \to & l^2(\N)\times l^2(\N)\\
         & U(t,k_2) v_j & \mapsto & (\delta_j,0)\\
         & U(t,k_2) T v_j & \mapsto & (0,\delta_j)
    \end{matrix}  
    \quad \mbox{where }\delta_j(n) = \begin{cases}
        1 \quad \mbox{if }n=j, \\
        0 \quad \mbox{otherwise.}
    \end{cases}\]
    This family enjoys the following properties: \[TI(t,k_2)^{-1} (\delta_j,0) = TU(t,k_2)v_j(0)= U(-t,-k_2)Tv_j(0) = I(-t,-k_2)^{-1}(0,\delta_j) \] \[TI(t,k_2)^{-1} (0,\delta_j) = TU(t,k_2)Tv_j= -U(-t,-k_2)v_j = I(-t,-k_2)^{-1}(-\delta_j,0) \] so by anti-linearity \[TI(t,k_2)^{-1} = I(-t,-k_2)^{-1} \K J \quad \mbox{where }J = \begin{pmatrix}
        0 & \Id_{l^2(\N)} \\ -\Id_{l^2(\N)} & 0
    \end{pmatrix} \quad \forall t\in[-\pi,\pi]\] and $\K$ is the standard complex conjugation.
    The lack of periodicity is expressed by the fact that \[I(-\pi,k_2)^{-1}I(\pi,k_2)\ne \Id_\Hi.\] 
    Using the spectral theorem, we know that there are two self-adjoint operators $L,R$ acting over $l^2(\N)^2$ with \[I(\pi,0) I(-\pi,0)^{-1}= e^{iL}\quad  \mbox{and}\quad I(\pi,\pi)I(-\pi,\pi)^{-1}=e^{iR}.\]
    Moreover, it is true that: \[\begin{split}\K J^{-1} I(\pi,0)I(-\pi,0)^{-1}& =I(-\pi,0) T^{-1} I(-\pi,0)^{1}=I(-\pi,0)I(\pi,0)^{-1}\K J^{-1}  \\ &= (I(\pi,0)I(-\pi,0)^{-1})^{-1} \K J^{-1}.\end{split}\] 
    This means that $\K J^{-1} e^{iL} \K J = e^{-iL}$, so $e^{-iJ^{-1}\overline{L}J}=e^{-iL}$. Therefore, $L$ can be taken such that $J^{-1}\overline{L} J= L$. We can do the same for $R$ to have that $J^{-1} \overline{R}J = R$. Using them, we can define the unitary operator $\beta(k_2)\in \U(l^2(\N)^2)$ with:  \[\beta (k_2) = \begin{cases}
        \exp\{\frac{i}{\pi}[(\pi-k_2)\frac{L}{2} + k_2 \frac{R}{2}]\} \quad &\mbox{if} \quad k_2\in[0,\pi]\\
        \K J^{-1} I(-\pi,-k_2) I(\pi,-k_2)^{-1}\beta(-k_2) \K J \quad&\mbox{if} \quad k_2\in [-\pi,0]
    \end{cases}\]
    and it is easily shown that $\beta(k_2)$ is continuous and periodic in $k_2$ by applying the symmetry relations. Now, thanks to Theorem \ref{thm:Kupier} we know that there is a homotopy $\beta_t(k_2)$ in $\U(l^2(\N)^2)$ with $\beta_0(k_2) \equiv \Id_{l^2(\N)^2}$ and $\beta_1(k_2)= \beta(k_2)$. To conclude the proof, we can define $\Tilde{I}^{-1}(t,k_2):l^2(\N)^2\to\ker(P(t,k_2))$ as: \[\Tilde{I}(t,k_2)^{-1} = \begin{cases}
         I(t,k_2)^{-1}\beta_{t/\pi}(k_2) & \mbox{for } t\in[0,\pi], k_2 \in\T^1 \\        I(t,k_2)^{-1}\K J^{-1} \beta_{-t/\pi}(-k_2) \K J  &\mbox{for }  t\in[-\pi,0], k_2 \in \T^1
    \end{cases}\]
    This $\Tilde{I}$ is continuous for $t=0$ because $\beta_0(k_2) \equiv \Id$, and is also periodic because, for $k_2 \in [0,\pi]$, we have that \[\begin{split} \Tilde{I}(-\pi,k_2) &= I(-\pi,k_2)^{-1} \K J^{-1} \beta_1 (-k_2) \K J  \\
    &= I(-\pi,k_2)^{-1}\K J^{-1} \K J^{-1} I(-\pi,k_2) I(\pi,k_2)^{-1} \beta (k_2) \K J\K J \\ & = I(\pi,k_2)^{-1} \beta(k_2) = \Tilde{I} (\pi,k_2) \end{split}\]
    while for $k_2 \in [-\pi,0]$ we have that \[
    \begin{split} \Tilde{I}(-\pi,k_2) &= I(-\pi,k_2)^{-1} \K J^{-1} \beta_1 (-k_2) \K J  \\
    &=I(\pi,k_2)^{-1} I(\pi,k_2) I(-\pi,k_2)^{-1}\K J^{-1} \beta (-k_2) \K J \\
    &=I(\pi,k_2)^{-1} \K J ^{-1} I(-\pi,-k_2) I(\pi,-k_2)^{-1} \beta(-k_2) \K J \\ & = I(\pi,k_2)^{-1} \beta(k_2) = \Tilde{I} (\pi,k_2) \end{split}
    \]
    Moreover, $\Tilde{I}$ also satisfies a symmetry constraint: \[\begin{split}T  \Tilde{I} (t,k_2)^{-1} & = \begin{cases}
        I(-t,-k_2)^{-1} \K J \beta_{t/\pi}(k_2) & \mbox{for }t\in[0,\pi]\\
        I(-t,-k_2)^{-1} \K J \K J^{-1} \beta_{-t/\pi}(-k_2) \K J & \mbox{for }t\in [-\pi,0]
    \end{cases}\\ &= \begin{cases}
        I(-t,-k_2)^{-1} \K J^{-1} \beta_{t/\pi}(k_2) \K J \K J & \mbox{for }t\in[0,\pi]\\
        I(-t,-k_2)^{-1}  \beta_{-t/\pi}(-k_2) \K J & \mbox{for }t\in [-\pi,0]
    \end{cases}\\
    &= \Tilde{I}(-t,-k_2)^{-1} \K J
    \end{split}\]

    Then $\{\Tilde{v}_j(t,k_2)=\Tilde{I}(t,k_2)^{-1} (\delta_j ,0)\}_{j\in \N}$ is the collection of vectors we were looking for, as it is periodic and continuous because so is $\Tilde{I}$. Moreover, elements such as $(0,\delta_j)$ together with $(\delta_j,0)$ form an orthonormal basis of $l^2(\N)^2$, so their images $\{\Tilde{v}_j(t,k_2) =\Tilde{I}(t,k_2)^{-1}(\delta_j,0)\}_{j\in \N}$ together with $\{\Tilde{u}_j(t,k_2) = \Tilde{I}(t,k_2)^{-1} (0,\delta_j)\}_{j\in\N}$ form a discrete orthonormal basis of $\ker P(t,k_2)$. Most importantly, the relation $(\delta_j,0) = \K J(0,\delta_j)$ implies that $\Tilde{v}_j (t,k_2) = T \Tilde{u}_j (-t,-k_2)$ for all $j \in \N, t\in [-\pi,\pi] , k_2\in \T^1$. So, the vectors \[\{\Tilde{v}_1(t), \Tilde{u}_1(t), \cdots, \Tilde{v}_j(t), \Tilde{u}_j(t), \cdots \}\] are exactly the collection of orthonormal vectors we needed to complete the proof.
\end{proof}

\begin{theorem}[Emerging spin-type symmetry] \label{thm:emerging spin}
Let $P:\T^2 \to \Proj_{2n}(\mathcal H)$ be a time-reversal symmetric projection-valued map. Then:
\begin{enumerate}
\item there exists a spin-type symmetry, namely a continuous unitary-valued map $S:\T^2 \to \mathcal U(\mathcal H)$ such that
\[
S(k)^2=\Id,\qquad [S(k),P(k)]=0,\qquad T S(k) = -S(-k) T,\qquad \forall k \in \T^2;
\]
\item the $\mathbb{Z}_2$ invariant of $P$ can be computed as a Spin-Chern number with respect to any spin-type symmetry $S$:
\[
\delta(P) = (-1)^{\Ch_S(P)}.
\]
\end{enumerate}
\end{theorem}

\begin{proof}
The existence of a map $S$ as in~(1) is equivalent to the existence of symmetric splittings of both $P$ and its orthogonal complement $P^\perp := \Id - P$
\[
P(k) = P^+(k) + P^-(k), \qquad P^\perp(k) = P^{\perp,+}(k) + P^{\perp,-}(k).
\]
Indeed, given such splittings, the operator
\[
S(k):=\Id - 2\big(P^+(k)+P^{\perp,+}(k)\big)
\]
satisfies $S(k)^2=\Id$, since $P^+(k)$ and $P^{\perp, +}(k)$ are orthogonal, hence $P^+(k)+P^{\perp,+}(k)$ is a projection. Moreover, $[S(k),P(k)]=0$, since both $P^+(k)$ and $P^{\perp,+}(k)$ commute with $P(k)$. Finally,
\[
\begin{aligned}
T S(k)
&= T\big(\Id - 2P^+(k)-2P^{\perp,+}(k)\big) \\
&= \big(\Id - 2P^-(-k)-2P^{\perp,-}(-k)\big)T = -S(-k)T,
\end{aligned}
\]
where the last equality follows from the identity
\[
P^+(k)+P^{\perp,+}(k) + P^-(k)+P^{\perp,-}(k)=\Id.
\]

A symmetric splitting of $P$ can always be obtained by means of Theorem~\ref{thm:equivalence_theorem}. The existence of such a splitting for $P^\perp$ follows from the same result when $\mathcal H$ is finite-dimensional, since in this case $P^\perp$ is a finite-rank time-reversal symmetric projection-valued map.

In the infinite-dimensional case, Lemma~\ref{lem:infinite_dim_Class_AII_frame_d=2} guarantees the existence of a family $\{v_i\}_{i\in\mathbb N}$ of continuous vector-valued functions such that $\{v_i\}_{i\in\mathbb N}\sqcup\{T v_i\}_{i\in\mathbb N}$ defines a frame for $P^\perp$. One then obtains the desired symmetric splitting of $P^\perp$ by defining
\[
P^{\perp,+}(k):=\sum_{i=0}^{\infty} |v_i(k)\rangle \langle v_i(k)|,\qquad
P^{\perp,-}(k):=\sum_{i=0}^{\infty} |T v_i(k)\rangle \langle T v_i(k)|.
\]

Statement~(2) follows directly from formula~\eqref{eq:Spin-Chern formula} and from Theorem~\ref{thm:equivalence_theorem}. Indeed, any spin-type symmetry $S$ induces a symmetric splitting $P = P^+ + P^-$ by setting $P^\pm := P (S \pm \Id)/2$, and therefore
\[
\delta(P)=(-1)^{\Ch(P^\pm)} = (-1)^{\Ch_S(P)}.\qedhere
\]
\end{proof}

\subsection{Completeness of the invariant}\label{subsec: Completeness of the invariant}

In Definition \ref{def:Equivalence relations for time-reversal symmetric projection-valued maps} we introduced a notion of Murray–von Neumann equivalence, a notion of symmetric unitary equivalence, and a notion of symmetric homotopy for time-reversal symmetric projectoion-valued maps. The goal of this section is to prove that $\delta$ is the only topological information needed to connect two different projection-valued maps by means of a symmetric homotopy. This is sufficient to show the completeness of the invariant with respect to all three equivalence relations, since homotopy equivalence is the weakest among them (see Remark \ref{rmk:Varie equivalenze}).

This result is partly known for other formulations of the invariant: for instance, completeness with respect to partial isometries (Murray–von Neumann equivalence) is discussed in the last section of~\cite{fiorenza2016z}, while in~\cite{Cornean_2017} completeness with respect to homotopy is discussed for the family of matching matrices, although an explicit homotopy between the associated projection-valued maps is not constructed — precisely what we will instead carry out in the following proof.

\begin{theorem}\label{thm:delta_complete_topological_invariant}
    Two time-reversal symmetric projection-valued maps $P_0,P_1:\T^2 \to \Proj_{2n}(\Hi)$ are symmetrically Murray--von Neumann equivalent, unitarily equivalent and homotopic if and only if $\delta(P_0)=\delta(P_1)$.
\end{theorem}

\begin{proof}
        We already know one part of the statement: in fact, Proposition \ref{pro:delta_unitarily_homotopic_invariant} states that if two maps are Murray--von Neumann equivalent, unitarily equivalent or homotopic, then their $\delta$ invariants are equal. 
        So, we want to construct a symmetric unitary equivalence between $P_0$ and $P_1$ under the assumption that $\delta(P_0)= \delta(P_1)$; the restriction of this unitary to the range of $P_0$ would then yield a Murray--von Neumann equivalence. We can start by applying Theorem \ref{thm:equivalence_theorem} to construct two pseudo-periodic symmetric frames \[\{v_1^0(k) , \cdots , v_n^0(k) , -Tv_1^0(-k), \cdots, -Tv_n^0(-k)\}\] \[\{v_1^1(k) , \cdots , v_n^1(k) , -Tv_1^1(-k), \cdots, -Tv_n^1(-k)\}\] of $P_0(k)$ and $P_1(k)$ respectively, such that \[\begin{matrix}
            v_1^0(\pi,k_2)= e^{i\pi h(\delta)k_2} v_1^0(-\pi,k_2) \\v_1^1(\pi,k_2)= e^{i\pi h(\delta)k_2} v_1^1(-\pi,k_2)
        \end{matrix}\quad \mbox{where }h(\delta) =\begin{cases}
            0 & \mbox{if } \delta(P_0)=\delta(P_1)=1\\
            1 & \mbox{if } \delta(P_0)=\delta(P_1) =-1
        \end{cases}\]
        Now we need to look at $\Id -P_0(k)$ and $\Id -P_1(k)$. These are once again two time-reversal symmetric projection-valued maps. If $\dim(\Hi)<\infty$, thanks to lemma \ref{lem:delta_additive}, it holds that \[\delta(\Id-P_0)= \delta(\Id)\cdot \delta(P_0) = \delta(P_0) = \delta(P_1) = \delta(\Id)\cdot \delta(P_1) = \delta(\Id-P_1).\]
        So we can apply Theorem \ref{thm:equivalence_theorem} to extend the previous frames obtaining:\[\{v_1^0(k) , \cdots, -Tv_n^0(-k), v_{n+1}^0(k) , \cdots, v_{\dim(\Hi)/2}^0(k), -Tv_{n+1}^0(-k) , \cdots -Tv_{\dim(\Hi)/2}^1(-k)\}\]\[\{v_1^1(k) , \cdots, -Tv_n^1(-k), v_{n+1}^1(k) , \cdots, v_{\dim(\Hi)/2}^1(k), -Tv_{n+1}^1(-k) , \cdots -Tv_{\dim(\Hi)/2}^1(-k)\}.\]
        where the added vectors span respectively $\ker P_0(k)$ and $\ker P_1(k)$ and have full periodicity with exceptions \[v_{n+1}^0(\pi,k_2) =e^{i\pi h(\delta)k_2} v_{n+1}^0(-\pi,k_2), \quad v_{n+1}^1(\pi,k_2)= e^{i\pi h(\delta) k_2} v_{n+1}^1(-\pi,k_2). \]
        Instead, if $\dim(\Hi)=\infty$ we can use the lemma \ref{lem:infinite_dim_Class_AII_frame_d=2} to extend the previous frames to \[\{v_1^0(k) , \cdots, -Tv_n^0(-k), v_{n+1}^0(k) , -Tv_{n+1}^0(-k) , \cdots, v_{N}^0(k),-Tv_{N}^0(-k), \cdots \}\]
        \[\{v_1^1(k) , \cdots, -Tv_n^1(-k), v_{n+1}^1(k) , -Tv_{n+1}^1(-k) , \cdots, v_{N}^1(k),-Tv_{N}^1(-k), \cdots \}\]
        where the added vectors have full periodicity and are two orthonormal bases of $\ker P_0(k)$ and $\ker P_1(k)$, respectively.
        
        In both cases, the unitary-valued map $V(k)$ we are looking for is the one such that $V(k)v_j^0(k) = v_j^1(k)$ and $V(k)Tv_j^0(-k)=Tv_j^1(-k) $ for all $j\in\{1,\cdots, \dim(\Hi)/2\}$ and $k \in \T^2$. In fact, it is trivial to prove that $P_1(k) = V(k)P_0(k)V(k)^{-1}$ and that $TV(k)=V(-k)T$. It is a bit less trivial to prove that this map is periodic despite the fact that it is defined over a non-periodic orthonormal basis. However, we can check, for example, that: \[\begin{split}v_1^1(\pi,k_2)&=V(\pi,k_2)v_1^0(\pi,k_2) =V(\pi,k_2)e^{ih(\delta)k_2} v_1^0(-\pi,k_2)=e^{ih(\delta)k_2}V(\pi,k_2) v_1^0(-\pi,k_2) \\
        &= e^{ih(\delta)k_2} v_1^1(-\pi,k_2) = e^{ih(\delta)k_2}V(-\pi,k_2)v_1^0(-\pi,k_2)\end{split} \] 
        so we have $V(-\pi,k_2)v_1^0(-\pi,k_2) = V(\pi,k_2)v_1^0(\pi,k_2)$. Similar arguments can be made for all non-periodic vectors to prove that $V(-\pi,k_2)$ acts as $V(\pi,k_2)$.
        
        Now we pass to consider homotopy equivalence, so if $\delta(P_0)=\delta(P_1)$, we want to exhibit a symmetric homotopy between the two projection-valued maps.
        By the previous discussion we can create a symmetric unitary equivalence $P_1(k) = V(k) P_0(k) V(k)^{-1}$ with $TV(k) = V(-k)T$. The first step is to define a unitary-valued map $V_0:\T^2 \to \U(\Hi)$ such that $TV_0(k) = V_0(-k)T$ and $P_0(k) = V_0(k) P_0(k) V_0(k)^{-1}$. If $\dim(\Hi) = \infty$ we can define $V_0(k) \equiv \Id$. Instead, if $\dim(\Hi) =2N <\infty$, we need to do something more. First, we need to use lemma \ref{lem:anti-unitary_quaternionic_structure} to select a basis on which $T$ acts as $J\K$. Then we need to observe that the condition $TV(k)=V(-k)T$ means that $J\overline{V(k)} = V(-k)J$. Set $\lambda(k) :=\det V(k)$, and consider the triangle $\triangle$ inside $[-\pi,\pi]^2$ with vertices \[Q_1=(-\pi,\pi),\quad Q_2=(\pi,\pi),\quad Q_3=(\pi,-\pi).\]  Then on the points $k_\star$ fixed by the involution $k\mapsto -k$ we have that $J= V(k_\star)^t J V(k_\star)$, so $V(k_\star) \in \rSp(2n)$. Moreover, on the segments $\overline{Q_1 Q_2}, \overline{Q_1,Q_3}, \overline{Q_2 Q_3}$, we have that $\lambda(k) = \overline{\lambda(-k)}$. Therefore, on those segments, we can apply Lemma \ref{lem:even_winding_number} to have that \[[\lambda(t,\pi)]_{t\in[-\pi,\pi] }=2m,\quad [\lambda(\pi,t)]_{t\in[-\pi,-\pi]}=2l,\quad [\lambda(t,-t)]_{t\in[-\pi,\pi]}=2h.\] Since the map is defined also inside the triangle, the winding number of $\lambda$ in $\triangle$ must be zero, so $2m = 2l+2h$. Now we can use the pseudo-periodic frame $\{v_1^0(k),\cdots, -Tv_N^0(-k)\}$ to define the unitary operator $V_0(k)$ with: \[V_0(k)v_j^0(k) =\begin{cases}
            e^{i(mk_1+lk_2)}v_1^0(k) &\mbox{if}\quad j=1\\
            v_j^0(k) & \mbox{otherwise}
        \end{cases} \]
        \[
            V_0(k)(-Tv_j^0(-k)) =\begin{cases}
            e^{i(mk_1+lk_2)}(-Tv_1^0(-k)) &\mbox{if}\quad j=1\\
            -Tv_j^0(-k) & \mbox{otherwise}
        \end{cases}\]
        This is actually periodic in $k$ because, for example, \[\begin{split}V_0(-\pi,k_2) v_1^0(\pi,k_2) &= V_0(-\pi,k_2) e^{i\pi h(\delta)} v_1^0(-\pi,k_2) = e^{i\pi h(\delta)}e^{i(-m\pi+lk_2)}v_1^0(-\pi,k_2)  \\ &=e^{-2im\pi}e^{i(m\pi+lk_2)} e^{i\pi h(\delta)}v_1^0(-\pi,k_2) = e^{i(m\pi+lk_2)} v_1^0(\pi,k_2) \\
        &=V_0(\pi,k_2)v_1^0(\pi,k_2) \end{split}\]
        In both cases $\{\dim(\Hi) = \infty, \dim(\Hi) = 2N\}$ we want to define a homotopy $V_t$ between $V_0$ and $V$. To do so, we need to work on $\triangle$. Notice that on the sides of the triangle $\overline{Q_1Q_2}$, $\overline{Q_1Q_3}$ and $\overline{Q_2Q_3}$ the projection-valued map behaves as one-dimensional time-reversal symmetric projection-valued maps:\[TP(k_1,\pi)= P(-k_1,-\pi)T = P(-k_1,\pi) T,\]\[TP(\pi,k_2)= P(-\pi,-k_2)T = P(\pi,-k_2)T,\] \[TP(k_1,-k_1)=P(-k_1,k_1)T.\] So we can use Theorem \ref{pro:Kato-Nagy,d=1} to build $V_t(k)$ for $k\in \partial\triangle$ with $TV_t(k) = V_t(-k)T$ for all $t\in[0,1], k\in\partial \triangle$. This means that $V_t(k)$ is defined for $(t,k) \in \partial([0,1]\times\triangle)$. This region however, is homeomorphic to $S^2$ and since $\pi_2(U(2N))= \{0\} = \pi_2(\U(\Hi))$, thanks to Theorems \ref{thm:homotopy_group_U(n)} and \ref{thm:Kupier}, it is always possible to extend the definition inside for $(t,k)\in[0,1]\times \triangle$. As a final adjustment, we just need to define the map in the complementary triangle by imposing $V_t(k) = T^{-1}V_t(-k)T$ for $k\in [-\pi,\pi]^2\setminus \triangle$. Then $V_t(k)$ is continuous, periodic and symmetric and therefore:\[P_t(k) =V_t(k) P_0(k) V_t(k)^{-1}\] is the symmetric homotopy we were looking for since $[P_0(k),V_0(k)]=0$. 
\end{proof}

\appendix

\section{Tools from algebraic geometry}\label{section: Tools from algebraic geometry}
In this appendix, we want to prove and highlight some aspects of quaternionic Hilbert spaces and winding numbers that we used in the main text of the paper. 
\begin{lemma}\label{lem:anti-unitary_quaternionic_structure}
    Given a complex Hilbert space $\Hi$ and an anti-unitary operator $T$ acting on $\Hi$ with $T^2=-\Id$, we can interpret the pair $(\Hi,T)$ as a quaternionic Hilbert space $\Hi_\mathbb{H}=(\Hi,T)$. Moreover, if $V\subset\Hi_\mathbb{H}$ is a quaternionic Hilbert subspace, namely a complex Hilbert subspace with $TV=V$, then the following statements are true:
    \begin{enumerate}
        \item $V$ admits a quaternionic basis.
        \item As a complex Hilbert subspace, $\dim(V)$ is infinite or even.
        \item If $\dim(V)=2L$ as a complex Hilbert space, it is always possible to build a complex basis of the form \[\{v_1,\cdots, v_{\dim(V)/2}, -Tv_1,\cdots ,- Tv_{\dim(V)/2}\}.\]
        This means that on this basis $T$ acts as $\K J$, for $\K$ standard complex conjugation and 
        \begin{equation} \label{eqn:J}
            J = \begin{pmatrix}         0&\id_n \\- \id_n & 0\end{pmatrix}.
        \end{equation}
    \end{enumerate} 
\end{lemma}

This argument is already present in several works, for example in~\cite{gontier2022symmetric}. We prove it below for the sake of completeness.

\begin{proof}
    We can define the set composed of the collection of orthonormal vectors that are orthogonal to their partners multiplied with $T$. Namely, we consider the set: \[Z= \left\{\{v_a\}_{a\in J}\subset \Hi |\inn{v_a}{v_b}=\delta_{b,a}, \inn{v_a}{Tv_b}=0 \ \forall a,b\in J \right\}\]
    This set has a natural partial order given by the inclusion, and every increasing sequence \[\{v_j\}_{j\in J_1}\subset\cdots \subset \{v_j\}_{j\in J_n}\subset \cdots \]
    can be viewed up to changing the indexes as a sequence $J_1\subset \cdots\subset J_n\subset\cdots$. Then we can consider $J=\cup_{i=1}^\infty J_i$ and look at $\{v_j\}_{j\in J}$, which is clearly an element of $Z$. The fact that $T^2=-\Id$ implies that \[\inn{v}{Tv}=\overline{\inn{Tv}{T^2 v}} = -\overline{\inn{Tv}{v}} = -\inn{v}{Tv}\] so $Z$ is not empty because the set consisting of a single normalized vector is an element of it, and we can apply the Zorn lemma to obtain that there is also a maximal element $\{v_j\}_{j\in J}$. This maximal element is such that \[\Hi=\Span_\C \left( \{v_j\}_{j\in\{1,\cdots, \dim(\Hi)/2\}}\right)\oplus \Span_\C \left( \{Tv_j\}_{j\in\{1,\cdots, \dim(\Hi)/2\}}\right).\]
    In fact, by definition, the two spans intersect only at $0$ and if, by contradiction, there is a vector $v$ that cannot be obtained by adding elements of the two spans, then, up to renormalization, the vector \[\Tilde{v} = v-\sum_{j\in J}\inn{v_j}{v} v_j - \sum_{j\in J} \inn{Tv_j}{v} Tv_j\] is non-zero and orthogonal to both spans. Moreover: \[\begin{split}T\Tilde{v} & =  Tv-\sum_{j\in J}\overline{\inn{v_j}{v} Tv_j} + \sum_{j\in J} \overline{\inn{Tv_j}{v}} v_j  \\
    &=Tv - \sum_{j\in J}\inn{Tv_j}{Tv} Tv_j - \sum_{j\in J} \inn{v_j}{Tv} v_j\end{split} \]
    So also $Tv$ is non-trivial and orthogonal to both spans, meaning that we can add $v$ to the collection $\{v_j\}_{j\in J}$ and contradict maximality.
\end{proof}

Next, we recall the definition and the basic properties of the winding number.

\begin{definition}[Winding number]\label{def:winding_number}
    Given a map $f:\rU(1) \to \rU(1) $, the \emph{winding number}, denoted $[f]$, is an integer number exhibited by the natural isomorphism $\pi_1(\rU(1) )\simeq \Z$. It can be computed by considering the natural covering $g:\R \to \rU(1) $ such that $g (t) = e^{i2\pi t}$. Then, the map $f$ can be lifted to a continuous map $\mu:[-\pi,\pi] \to \R$ such that $g(\mu(t))=f(\pi(t))$. Despite the numerous lifting options, the quantity $[f]=\mu(\pi)-\mu(-\pi) \in \Z$ is well defined and depends only on the homotopy class of $f$.
\end{definition}

\begin{lemma}\label{lem:even_winding_number} 
    If $\lambda: \T^1\to \rU(1) $ is a continuous map with $\lambda(\pi)=\lambda(-\pi)=\lambda(0) = 1$ and such that $\lambda(-k) = \overline{\lambda(k)}$, then $[\lambda] \in 2\Z$ is even.
\end{lemma}

\begin{proof}
    We can choose a lift $\mu:[-\pi,\pi] \to \R$ such that $e^{i\mu(k)} = \lambda(k)$ with $\mu (0) = 0$. Then the properties in the statement translate to the fact that $\mu(k) = -\mu(-k)$ and $\mu(\pi),\mu(-\pi) \in 2\pi\Z$. So, the winding number will be:\[[\lambda]=\frac{\mu(\pi)-\mu(-\pi)}{2\pi}=\frac{2 \mu (\pi)}{2\pi} \in 2\Z. \qedhere\]
\end{proof}

\begin{lemma}\label{lem:triviality_of_reflective_map}
    If $\lambda:\T^1\to \T^1$ is a continuous map with $\lambda(k)=\lambda(-k)$, then $[\lambda]=0$. 
\end{lemma}

\begin{proof}
    We can construct an explicit homotopy $\lambda_t$ for $t\in[0,1]$ between $\lambda$ and the constant function $\lambda_0\equiv\lambda(\pi)=\lambda(-\pi)$: \[
    \lambda_t(k) = \begin{cases}
        \lambda(k)&\mbox{for}\quad k\in[-\pi,(t-1)\pi] \cup[(1-t)\pi,\pi]\\
        \lambda[(t-1)\pi]= \lambda[(1-t)\pi]&\mbox{for} \quad k\in[(t-1)\pi, (1-t)\pi]
    \end{cases}
    \]
    In fact, the two branches of the function connect continuously for $k \to (t-1)\pi$, $k \to (1-t)\pi$ and also for $k\to\pi$, $k \to -\pi$ since $\lambda(\pi)=\lambda(-\pi)$. This means that $\lambda_t(k)$ is continuous.
\end{proof}

\begin{proposition}\label{cor:top_group} 
Consider $\rU(1)  = \left\{ e^{i\theta} \in \mathbb{C}\right\} $ as a topological group with the standard product of $\mathbb{C}$. If $f,g : \rU(1)  \to \rU(1) $ are two smooth maps with $f(1) = g(1) = 1$ then the winding number of $fg $ is the sum of the winding number of $f$ and the one of $g$.
\end{proposition}
\begin{proof}
By \cite[Theorem~1.7]{hatcher2005algebraic}, the map
\[
\mathbb{Z}\ni n \longmapsto \big[\alpha \mapsto e^{i2\pi n \alpha}\big]\in \pi_1(\rU(1),1)
\]
is a group isomorphism. Moreover, if $f_0\simeq f_1$ and $g_0\simeq g_1$ are homotopies of based loops in $\rU(1)$, then the pointwise product $f_0 g_0$ is homotopic to $f_1 g_1$, since multiplication in $\rU(1)$ is continuous. Therefore, the assignment sending a loop to its winding number is compatible with pointwise multiplication of loops.

Applying this observation to the representatives $f$ and $g$, one concludes that the homotopy class of $f g$ corresponds to the sum of the homotopy classes of $f$ and $g$, and hence that the winding number of $fg$ is the sum of the winding numbers of $f$ and $g$.
\end{proof}

\begin{theorem}[{Homotopy groups of $\rU(n)$~\cite[Chapter 8, Section 12]{husemoller1966fibre}}]\label{thm:homotopy_group_U(n)}
    If $\Hi$ is a complex Hilbert space with finite dimension, then $\pi_0(\U(\Hi)) \simeq \{0\}$, $\pi_1(\U(\Hi)) \simeq \Z$ using the homomorphism that computes the winding number of the determinant, and $\pi_2(\U(\Hi))\simeq \{0\}$.
\end{theorem}
\begin{theorem}[{Kuiper's theorem~\cite{kuiper1965homotopy}}]\label{thm:Kupier}
    If $\Hi$ is a real, complex, or quaternionic separable Hilbert space with infinite dimension, then $\U (\Hi)$ is weakly contractible.
\end{theorem}

\bibliographystyle{plain}
\bibliography{bibliography}

\bigskip \bigskip

{\footnotesize

\begin{tabular}{ll}
(A.~Ferreri) & \textsc{Dipartimento di Matematica, Sapienza Universit\`{a} di Roma} \\
 &  Piazzale Aldo Moro 5, 00185 Rome, Italy \\
 &  {E-mail address}: \href{mailto:alessandro.ferreri@uniroma1.it}{\texttt{alessandro.ferreri@uniroma1.it}} \\[5pt]
(D.~Monaco) & \textsc{Dipartimento di Matematica, Sapienza Universit\`{a} di Roma} \\
 &  Piazzale Aldo Moro 5, 00185 Rome, Italy \\
 &  {E-mail address}: \href{mailto:domenico.monaco@uniroma1.it}{\texttt{domenico.monaco@uniroma1.it}} \\[5pt]
(G.~Peluso) & \textsc{Dipartimento di Matematica, Sapienza Universit\`{a} di Roma} \\
 &  Piazzale Aldo Moro 5, 00185 Rome, Italy \\
 &  {E-mail address}: \href{mailto:gabriele.peluso@uniroma1.it}{\texttt{gabriele.peluso@uniroma1.it}} \\
\end{tabular}
}

\end{document}